# Integrating Response Time and Attention Duration in Bayesian Preference Learning for Multiple Criteria Decision Aiding

Jiaxuan Jiang[a], Jiapeng Liu[a,*], Miłosz Kadziński[b], Xiuwu Liao[a,c], Jingyu Dong[d]

[a]*Center for Intelligent Decision-Making and Machine Learning, School of Management, Xi'an Jiaotong University, Xi'an, Shaanxi 710049, PR China*
[b]*Institute of Computing Science, Poznan University of Technology, Piotrowo 2, Poznań 60-965, Poland*
[c]*Collaborative Innovation Center of China Pilot Reform Exploration and Assessment - Hubei Sub-Center, Hubei University of Economics, Wuhan, Hubei 430205, PR China*
[d]*School of Management, Xi'an Jiaotong University, Xi'an, Shaanxi 710049, PR China*

## Abstract

We introduce a multiple criteria Bayesian preference learning framework incorporating behavioral cues for decision aiding. The framework integrates pairwise comparisons, response time, and attention duration to deepen insights into decision-making processes. The approach employs an additive value function model and utilizes a Bayesian framework to derive the posterior distribution of potential ranking models by defining the likelihood of observed preference data and specifying a prior on the preference structure. This distribution highlights each model's ability to reconstruct Decision-Makers' holistic pairwise comparisons. By leveraging both response time as a proxy for cognitive effort and alternative discriminability as well as attention duration as an indicator of criterion importance, the proposed model surpasses traditional methods by uncovering richer behavioral patterns. We report the results of a laboratory experiment on mobile phone contract selection involving 30 real subjects using a dedicated application with time-, eye-, and mouse-tracking components. We validate the novel method's ability to reconstruct complete preferences. The detailed ablation studies reveal time- and attention-related behavioral patterns, confirming that integrating comprehensive data leads to developing models that better align with the DM's actual preferences.

*Keywords:* Decision analysis, Behavioral data, Response time, Attention duration, Bayesian inference

## 1. Introduction

Multiple Criteria Decision Aiding (MCDA) provides a robust framework for addressing decision problems involving a finite set of alternatives evaluated across multiple criteria [14]. Over the past decades, these problems have been widely applied across diverse fields such as economics [4], engineering [2], education [5], environment [36], and medicine [31]. In this paper, we focus on multiple criteria choice and ranking problems [14], where deriving decision recommendations requires eliciting preference information from the Decision Maker (DM). An important trend in MCDA is using indirect preference information provided through examples of holistic judgments, such as pairwise comparisons or DMs' choices [29]. This approach is less cognitively demanding than the direct specification of the preference model's parameter values, forming the foundation of the ordinal regression paradigm [25, 35]. Ordinal regression constructs a preference model that aligns with the provided decision examples, allowing the evaluation of a set of alternatives and working out recommendations for a specific problem to solve.

---
*Corresponding author

*Email addresses:* `3120108031@stu.xjtu.edu.cn` (Jiaxuan Jiang), `jiapengliu@mail.xjtu.edu.cn` (Jiapeng Liu), `milosz.kadzinski@cs.put.poznan.pl` (Miłosz Kadziński), `liaoxiuwu@mail.xjtu.edu.cn` (Xiuwu Liao), `3120108096@stu.xjtu.edu.cn` (Jingyu Dong)

While pairwise comparisons and choices remain central to preference elicitation in MCDA, the decision-making process generates a wealth of auxiliary process data that often goes unutilized. Valuable behavioral data includes response time [19], attention patterns across information sources, eye-tracking metrics (e.g., fixations, saccades, and gaze duration), and mouse/cursor movements (e.g., visual scanning paths between alternatives or trajectories during interactions) [58]. Furthermore, expressed emotions [74] and physiological measures – such as heart rate variability, somatic markers, skin conductance, and neural activation data – offer objective insights into the cognitive processes that underlie decision-making [9, 26, 70]. Self-reported data – including confidence ratings, perceived difficulty, satisfaction, and mood assessments – provide a complementary subjective perspective, helping to capture the internal experiences of DMs [52, 67]. Moreover, analyzing error rates – such as inconsistencies in choice patterns or reversals – as detailed, e.g., in the diffusion decision model [69], reveals important information about the efficiency and potential biases in cognitive processing during choice tasks. In the same spirit, a swift decision may reflect strong preferences or significant differences between alternatives, while hesitation may indicate a lack of differentiation. Overall, such process data offers valuable insights into the decision mechanisms, shedding light on DMs' preferences in ways that traditional methods cannot [1, 22]. Despite their potential, these rich sources of information remain largely unexplored in the MCDA literature.

Nevertheless, recent research in behavioral and decision sciences has increasingly recognized the value of process data in understanding decision-making. The digitalization of experimental and transactional environments has significantly lowered the cost and complexity of collecting such data [23]. Among these, response time has emerged as a particularly promising indicator. Defined as the time required for a DM to make a decision, it offers a non-invasive, readily observable measure of the decision process [20, 23]. Clithero [22] further demonstrated that incorporating response time alongside choice outcomes enhances the predictive accuracy of value-based decision models. Additionally, shorter response times are often observed when alternatives differ substantially in perceived value, reflecting the discriminability of alternatives [11, 71]. Beyond response time, attention duration has been employed to examine how DMs allocate cognitive resources across different information cues. Some scholars have adopted time-based measures, considering the duration spent on specific attributes as an indicator of their relative importance [18, 33, 59]. Busemeyer and Townsend [17] further explored risk-based choices and suggested that DMs dynamically reallocate attention between attributes as their preferences evolve. Together, response time and attention duration offer complementary insights into decision processes, capturing the overall cognitive effort involved and the distribution of attention among competing attributes.

The theoretical results and experimental evidence underscore the importance of incorporating behavioral data into decision problems. Also, in a broader perspective, integrating human and behavioral aspects into real-world processes is essential for bridging the theories to the practice of operational research with a view to designing or improving methods, systems, and operations [30].

Building on this foundation, we make the following three significant contributions to the field of MCDA. First, we propose a novel preference learning model for multiple criteria choice and ranking problems that integrates response time and attention duration as distinct forms of indirect preference information. Traditional MCDA approaches primarily focus on explicit judgments of alternatives, overlooking the cognitive effort and deliberation underlying decision-making. By incorporating behavioral cues in relation to pairwise comparisons for a subset of reference alternatives, our study deepens the understanding of how DMs evaluate alternatives and prioritize criteria. Accounting for response time and attention duration, we gain insights into DMs' cognitive engagement, perceived task difficulty, and confidence levels. This enriched perspective enables a more nuanced interpretation of preferences and uncovers implicit preferences that conventional models tend to overlook. Ultimately, by leveraging this preference information, our approach enhances the accuracy of preference modeling and offers a more comprehensive view of individual decision behavior.

Our second key contribution is developing a Bayesian preference learning method that integrates multiple types of preference information – pairwise comparisons, response time, and attention duration – into a unified framework. We adopt an additive value function as the DM's preference model and construct its posterior distribution within a Bayesian framework, reflecting the inherent uncertainty in the elicited preference information. Specifically, the additive value function is treated as a random variable, and its posterior distribution is derived by modeling the joint likelihood of observed multiple types of indirect preference information, given a specific value function model. Such a likelihood-based methodology differentiates all potential value functions according to their ability to generate the observed data, ensuring that the posterior distribution remains centered on the DM's true preference model. This way, the framework constructs a structured representation of the DM's underlying preferences, while systematically capturing the intricate relationships among three diverse information sources. Overall, this Bayesian approach provides a rigorous foundation for preference inference, ultimately improving the robustness and reliability of decision analysis.

Finally, this research deepens the understanding of the role of behavioral cues in MCDA by examining their relationship with individual decision patterns and task characteristics. We conducted a laboratory experiment on contract selection to demonstrate the applicability and robustness of the proposed model. During the tests involving 30 real subjects, we systematically collected data on attention allocation to individual criterion, response time for each pairwise comparison, and mouse cursor movement patterns, enabling a detailed analysis of decision-making behavior. We report the results of ablation studies and comparative analyses involving the proposed method, its variant models, including only attention to criteria, those incorporating only response time, and the Bayesian ordinal regression model introduced by [63]. Furthermore, we comprehensively analyze participants' decision-making patterns across tasks and criteria. This lets us uncover insights into the role of behavioral cues in understanding individual preferences. In particular, we focus on investigating the role of response time as a proxy for cognitive effort and the discriminability of alternatives. We also study the capacity of attention duration to reveal the relative importance assigned to criteria during the decision process. We demonstrate that together, these elements provide a richer and more comprehensive representation of individual preferences, enhancing both the interpretability and predictive accuracy of the preference model. Above all, the insights gained from the study deepen our understanding of decision behavior and offer guidance for personalized interventions.

The remainder of the paper is organized in the following way. Section 2 is devoted to the problem formulation, allowed preference information, and employed preference model. Section 3 reviews the experiments involving real-world subjects in MCDA. Section 4 illustrates the applicability of the proposed model to a laboratory experiment on contract selection and demonstrates the performance and robustness of the approach through comparative experiments. Section 5 concludes the paper and provides avenues for future research.

## 2. Preference learning framework accounting for response time and attention duration in preference elicitation

*2.1. Problem description*

Let us consider a multiple criteria decision problem where the DM aims to rank a finite set of alternatives from the best to the worst or choose the most preferred one. Formally, we denote a set $\mathcal{A}$ of $N$ alternatives, $\mathcal{A} = \{a_1, \ldots, a_n, \ldots, a_N\}$, and assume each alternative is evaluated based on a family of $M$ criteria $\mathcal{G} = \{g_1, \ldots, g_m, \ldots, g_M\}$. Each criterion $g_m : \mathcal{A} \to \mathbb{R}$ maps an alternative $a_n$ to its performance score $g_m(a_n)$ on criterion $g_m$. Without loss of generality, we assume that all criteria are of gain type, meaning that a larger value of $g_m(a_n)$ indicates a more preferred performance of alternative $a_n$ on criterion $g_m$. We further postulate that the DM's preferences can be described using a preference model $U(\cdot)$, which aggregates the performances of each alternative $a_n \in \mathcal{A}$ on all the criteria $g_m \in \mathcal{G}$ to derive a comprehensive numerical score $U(a_n)$. In this paper, we use the following additive value function [46] as the preference model to compute $U(a_n)$ for each alternative $a_n$:

$$U(a_n) = \sum_{m=1}^{M} U_m(g_m(a_n)), \tag{1}$$

where $U_m(g_m(a_n))$ is a marginal value of alternative $a_n$ on criterion $g_m$.

To construct the preference model $U(\cdot)$, the prevailing methods in MCDA usually involve eliciting indirect preference information from the DM. Such information typically consists of holistic pairwise comparisons based on a preference relation $\succ$ defined over a subset of so-called reference alternatives $\mathcal{A}^R \subseteq \mathcal{A}$. Existing MCDA methods only consider such holistic judgments using the preference-disaggregation paradigm for calibrating the constructed preference model. This paper considers additional information, including response time and attention duration, in the preference elicitation procedure. We aim to extract extra preference cues from such auxiliary information to construct the preference model. Specifically, suppose that we collect from the DM $L$ pieces of preference information, denoted as $\mathcal{Q} = \{q_1, \ldots, q_l, \ldots, q_L\}$, and each preference information $q_l$, $l = 1, \ldots, L$, comprises three components $(I_l, R_l, \{T_{lm}\}_{m=1}^{M})$:

- **Pairwise Comparison:** $I_l$ specifies a preference relation between two reference alternatives $(a_{s_1^l}, a_{s_2^l}) \in \mathcal{A}^R \times \mathcal{A}^R$, represented by $a_{s_1^l} \succ a_{s_2^l}$ indicating that $a_{s_1^l}$ is preferred to $a_{s_2^l}$. The set of all pairwise comparisons is denoted by $\mathcal{I} = \{I_1, \ldots, I_l, \ldots, I_L\}$.

- **Response Time:** $R_l$ represents the total time taken by the DM in specifying the $l$-th pairwise comparison $I_l$, including both the processing time and the decision-making time. The set of response times is denoted by $\mathcal{R} = \{R_1, \ldots, R_l, \ldots, R_L\}$.

- **Attention Duration:** $T_{lm}$ represents the duration of attention allocated to criterion $g_m$ during the $l$-th pairwise comparison $I_l$. The set $\{T_{lm}\}_{m=1}^{M}$ captures the respective focus assigned to each criterion, providing fine-grained cues about the DM's decision-making process. The set of attention durations is denoted by $\mathcal{T} = \{T_{11}, \ldots, T_{lm}, \ldots, T_{LM}\}$.

In our implementation, the various types of preference information mentioned above are recorded using a *Tobii eye tracker*. Pairwise comparison data are obtained from the mouse click positions, while response time is indicated by the duration spent on each task. Attention duration is processed by treating each criterion as a Region of Interest (ROI), which is delineated using a rectangular box. The gaze time on an ROI is defined as the sum of the durations of all fixations that land on that ROI, and this is also recorded using the eye tracker. Detailed information is provided in Section 4.1.

Our objective is to construct a preference model $U(\cdot)$ based on the composite preference information $\mathcal{Q}$. The preference model will be established based on the pairwise comparisons $\mathcal{I}$, the respective response times $\mathcal{R}$, and attention durations $\mathcal{T}$.

*2.2. Model Formulation*

In this section, we propose a novel preference learning approach that jointly considers pairwise comparisons, response times, and attention durations within a Bayesian framework. We assume that the DM owns an inherent preference model $U(\cdot)$ and provides preference information according to this model. We aim to estimate the DM's underlying preference model based on the observed preference information using the Bayesian inference paradigm. Specifically, we derive the posterior distribution $P(U \mid \mathcal{Q})$ of the value function model $U(\cdot)$ using the following Bayes' rule:

$$P(U \mid \mathcal{Q}) = \frac{P(U)P(\mathcal{I}, \mathcal{R}, \mathcal{T} \mid U)}{P(\mathcal{I}, \mathcal{R}, \mathcal{T})} = \frac{P(U)P(\mathcal{I} \mid U)P(\mathcal{R} \mid U)P(\mathcal{T} \mid U)}{P(\mathcal{I}, \mathcal{R}, \mathcal{T})}, \quad (2)$$

where $P(\mathcal{I}, \mathcal{R}, \mathcal{T} \mid U)$ represents the likelihood of the pairwise comparisons $\mathcal{I}$ and additional preference cues $\mathcal{T}$ and $\mathcal{R}$, $P(U)$ is the prior distribution of the value function model $U(\cdot)$, and $P(\mathcal{I}, \mathcal{R}, \mathcal{T})$ is a normalization term. In particular, the likelihood function $P(\mathcal{I}, \mathcal{R}, \mathcal{T} \mid U)$ is decomposed into three components $P(\mathcal{I} \mid U)$, $P(\mathcal{R} \mid U)$, and $P(\mathcal{T} \mid U)$, by assuming that $\mathcal{I}$, $\mathcal{R}$, and $\mathcal{T}$ are mutually independent given the preference model $U(\cdot)$. We adopt this decomposition due to the following reasons. First, it breaks the composite likelihood $P(\mathcal{I}, \mathcal{R}, \mathcal{T} \mid U)$ into manageable parts, each of which can be specified separately in an easier way. Hence, it enhances the model's flexibility in interpreting different types of preference information. Second, it requires a relatively less amount of preference information elicited from the DM for inferring the parameters of the value function model $U(\cdot)$, as we do not need to model the joint conditional distribution $P(\mathcal{I}, \mathcal{R}, \mathcal{T} \mid U)$, which resides in a high-dimensional space. This is analogous to the consideration in the famous naive Bayes classifier, which also incorporates the conditional independence assumption to avoid the curse of dimensionality [61]. Note that the realism of such an assumption is underscored by low numbers of pairwise comparisons in typical MCDA applications, where indirect preference information is elicited from the DMs rather than imposed by historical data [25].

According to the above Bayes rule, it can be observed that the posterior distribution of the preference model integrates impacts from both the likelihood function that compiles the DM's preference information and the prior information about the model's parameters. In subsequent sub-sections, we elaborate on how to define the likelihood terms $P(\mathcal{I} \mid U)$, $P(\mathcal{R} \mid U)$, and $P(\mathcal{T} \mid U)$ as well as the prior $P(U)$. Particularly, we discuss diverse model variants that account for different forms of response time and attention duration, which allows for designing tailored methods in modeling preference cues based on how these factors are treated.

*2.2.1. Defining the probability of generating each pairwise comparison*

We opt for the Bradley-Terry model [13] to characterize the probability of generating each pairwise comparison $I_l$, $l = 1, \ldots, L$, according to the underlying preference model $U(\cdot)$. By adopting an exponential form, the Bradley-Terry model constructs a distribution for each pairwise comparison, establishing a clear and intuitive relationship between the probability of the DM preferring one alternative to another and the difference in the comprehensive values for that pair of alternatives. Assuming independence among pairwise comparisons, the probability of generating all the pairwise comparisons, conditional on the preference model $U(\cdot)$, can be expressed as follows:

$$P(\mathcal{I} \mid U) = \prod_{l=1}^{L} P(I_l : a_{s_1^l} \succ a_{s_2^l} \mid U) = \prod_{l=1}^{L} \frac{\exp(U(a_{s_1^l}))}{\exp(U(a_{s_1^l})) + \exp(U(a_{s_2^l}))}, \quad (3)$$

where larger difference between $U(a_{s_1^l})$ and $U(a_{s_2^l})$ corresponds to higher likelihood of the DM asserting $a_{s_1^l} \succ a_{s_2^l}$. This is reasonable because, for additive representations of the considered preferences, the scale of the value function is a joint interval scale; consequently, the value difference has a meaningful interpretation of intensity. The Bradley-Terry model has been previously adopted in [63] for addressing multiple criteria ranking, where readers can find a more detailed justification.

*2.2.2. Modeling response time for pairwise comparisons*

As discussed previously, response time, representing the duration the DM spends evaluating reference alternatives, serves as cues of her preferences. By analyzing response time in pairwise comparisons, we can gain insights into how the DM allocates her attention and makes decisions. This information can then be used to construct the DM's preference model more accurately, leading to a more comprehensive understanding of the DM's preferences. In this section, we explore how to involve response time for each comparison in the proposed approach.

The underlying assumption consists in that the time the DM spends comparing a pair of reference alternatives $(a_{s_1^l}, a_{s_2^l}) \in \mathcal{A}^R \times \mathcal{A}^R$ depends on how easily they can be distinguished. Specifically, when the difference in values $U(a_{s_1^l})$ and $U(a_{s_2^l})$ is more significant, it is less demanding of the cognitive effort for the DM to make a comparison, thus leading to a shorter response time. Conversely, when $U(a_{s_1^l})$ and $U(a_{s_2^l})$ are closer, the DM may hesitate in making a judgment, resulting in a longer response time. To implement this assumption, we use an auxiliary variable $\delta(a_{s_1^l}, a_{s_2^l}) = \left|U(a_{s_1^l}) - U(a_{s_2^l})\right| \in [0,1]$ to quantify the absolute difference in values between alternatives $a_{s_1^l}$ and $a_{s_2^l}$. Then, we can explore diverse distributions to relate the value difference $\delta(a_{s_1^l}, a_{s_2^l})$ to the response time for the pairwise comparison $R_l$. In this paper, we consider the exponential, Gamma, and Poisson distributions as three alternative options:

- A standard approach for modeling the relationship between the value difference $\delta(a_{s_1^l}, a_{s_2^l})$ and the response time $R_l$ is to assume that $R_l$ follows an exponential distribution, i.e., $R_l \sim Exponential(\cdot \mid \lambda_l)$, in which the rate parameter $\lambda_l$ is regressed on the value difference $\delta(a_{s_1^l}, a_{s_2^l})$ by taking the following log-linear form:

$$\log \lambda_l = c_1 \delta(a_{s_1^l}, a_{s_2^l}) + c_2 = c_1 \left|U(a_{s_1^l}) - U(a_{s_2^l})\right| + c_2, \tag{4}$$

where $c_1$ and $c_2$ are regression parameters. The former ($c_1$) quantifies the DM's sensitivity to the value difference $\delta(a_{s_1^l}, a_{s_2^l})$, while the latter ($c_2$) adjusts the baseline level of response time, reflecting individual heterogeneity in processing speed. This form ensures that $\lambda_l > 0$ since $\lambda_l = \exp(c_1 \left|U(a_{s_1^l}) - U(a_{s_2^l})\right| + c_2)$. In the case of $c_1 > 0$, it verifies the assumption that the larger the value difference $\delta(a_{s_1^l}, a_{s_2^l})$, the less response time the DM spends in making a comparison between $a_{s_1^l}$ and $a_{s_2^l}$. This is because the value difference $\delta(a_{s_1^l}, a_{s_2^l})$ is positively related to the parameter $\lambda_l$, and the mean of $Exponential(\cdot \mid \lambda_l)$ is $\frac{1}{\lambda_l}$, meaning that a larger $\delta(a_{s_1^l}, a_{s_2^l})$ leads to a shorter response time $R_l$ when $c_1 > 0$. In turn, by assuming independence among pairwise comparisons, the likelihood of observing specific response time in pairwise comparisons, conditional on the preference model $U(\cdot)$, is formally described as follows:

$$P(\mathcal{R} \mid U) = \prod_{l=1}^{L} P(R_l \mid U) = \prod_{l=1}^{L} \lambda_l \exp(-\lambda_l R_l). \tag{5}$$

- Another approach to modeling response time is to use a Gamma distribution by assuming the response time $R_l \sim Gamma(\cdot \mid \alpha_l, \beta_l)$, where $\alpha_l > 0$ and $\beta_l > 0$ are the shape and rate parameters, respectively. Such a distribution is particularly suitable for capturing the variability inherent in the decision-making process. Parameters $\alpha_l$ and $\beta_l$ are regressed on the value difference $\delta(a_{s_1^l}, a_{s_2^l})$ by taking the following log-linear form:

$$\begin{aligned} \log \alpha_l &= c_1 \delta(a_{s_1^l}, a_{s_2^l}) + c_2 = c_1 \left|U(a_{s_1^l}) - U(a_{s_2^l})\right| + c_2, \\ \log \beta_l &= c_3 \delta(a_{s_1^l}, a_{s_2^l}) + c_4 = c_3 \left|U(a_{s_1^l}) - U(a_{s_2^l})\right| + c_4, \end{aligned} \tag{6}$$

where parameters $c_1$ and $c_3$ control how value differences impact the shape and scale of the distribution, and parameters $c_2$ and $c_4$ adjust the baseline levels of response time, reflecting individual heterogeneity in processing speed. In the case of $c_1 < 0$ and $c_3 < 0$, it verifies the assumption that the larger the value difference $\delta(a_{s_1^l}, a_{s_2^l})$, the less response time the DM spends in making a comparison between $a_{s_1^l}$ and $a_{s_2^l}$. This is because the value difference $\delta(a_{s_1^l}, a_{s_2^l})$ is negatively related to the parameters $\alpha_l$ and $\beta_l$, and the mean of $Gamma(\cdot \mid \alpha_l, \beta_l)$ is $\alpha_l \beta_l$, meaning that a larger $\delta(a_{s_1^l}, a_{s_2^l})$ derives shorter response time $R_l$ when $c_1 < 0$ and $c_3 < 0$. Analogously, by assuming independence among pairwise comparisons, the likelihood $P(\mathcal{R} \mid U)$ is formally described as follows:

$$P(\mathcal{R} \mid U) = \prod_{l=1}^{L} P(R_l \mid U) = \prod_{l=1}^{L} \frac{R_l^{\alpha_l - 1} \exp(-R_l / \exp(\beta_l))}{\beta_l^{\alpha_l} \Gamma(\alpha_l)}. \tag{7}$$

- An alternative approach to modeling response time is through a Poisson distribution, which is a discrete probability distribution expressing the probability of a given number of events. In our context, by rounding the response time $R_l$ up to the nearest integer $\lceil R_l \rceil$, we assume $\lceil R_l \rceil$ following a Poisson distribution, i.e., $\lceil R_l \rceil \sim Poisson(\cdot \mid \lambda_l)$, where the parameter $\lambda_l > 0$ is regressed on the value difference $\delta(a_{s_1^l}, a_{s_2^l})$ by taking the following log-linear form:

$$\log \lambda_l = c_1 \delta(a_{s_1^l}, a_{s_2^l}) + c_2 = c_1 \left| U(a_{s_1^l}) - U(a_{s_2^l}) \right| + c_2, \tag{8}$$

where parameter $c_1$ quantifies the DM's sensitivity to the value difference $\delta(a_{s_1^l}, a_{s_2^l})$, while parameter $c_2$ adjusts the baseline level of response time, reflecting individual heterogeneity in processing speed. In the case of $c_1 < 0$, it verifies the assumption that the larger the value difference $\delta(a_{s_1^l}, a_{s_2^l})$, the shorter response time the DM spends in making a comparison between $a_{s_1^l}$ and $a_{s_2^l}$. This is because the value difference $\delta(a_{s_1^l}, a_{s_2^l})$ is negatively related to the parameter $\lambda_l$, and the mean of $Poisson(\cdot \mid \lambda_l)$ is $\lambda_l$, meaning that a larger $\delta(a_{s_1^l}, a_{s_2^l})$ implies shorter response time $R_l$ when $c_1 < 0$. Then, by assuming independence among pairwise comparisons, the likelihood $P(\mathcal{R} \mid U)$ is formally described as follows:

$$P(\mathcal{R} \mid U) = \prod_{l=1}^{L} P(R_l \mid U) = \prod_{l=1}^{L} \frac{\lambda_l^{\lceil R_l \rceil} \exp(-\lambda_l)}{\lceil R_l \rceil!}. \tag{9}$$

*2.2.3. Modeling attention duration among criteria*

In addition to the response time for pairwise comparison, we can also model the attention duration for different criteria as extra cues for inferring the DM's preferences. Analogous to modeling response time in Section 2.2.2, we assume the DM's attention duration $T_{lm}$ for a particular criterion $g_m$ when making the pairwise comparison $I_l$ is negatively related to the difference between marginal values on criterion $g_m$ of the corresponding alternatives $a_{s_1^l}$ and $a_{s_2^l}$. That is, the larger the difference between marginal values $\delta_m(a_{s_1^l}, a_{s_2^l}) = \left| U_m(a_{s_1^l}) - U_m(a_{s_2^l}) \right|$, the lesser the DM's attention duration $T_{lm}$. Following the above assumption, we can use the exponential, Gamma, or Poisson distribution to model the attention duration $T_{lm}$ similarly to what is described in Section 2.2.2:

- The first option is to regress the attention duration $T_{lm}$ using an exponential distribution, i.e., $T_{lm} \sim Exponential(\cdot \mid \lambda_{lm})$, where the rate parameter $\lambda_{lm}$ is dependent on the difference $\delta_m(a_{s_1^l}, a_{s_2^l})$ by:

$$\log \lambda_{lm} = c_1 \delta_m(a_{s_1^l}, a_{s_2^l}) + c_2 = c_1 \left| U_m(a_{s_1^l}) - U_m(a_{s_2^l}) \right| + c_2, \tag{10}$$

where the interpretation of parameters $c_1$ and $c_2$ is the same as for Equation (4). Then, by assuming independence among criteria, the likelihood for all attention durations in $\mathcal{T}$ conditional on the preference model $U(\cdot)$ is formulated as follows:

$$P(\mathcal{T} \mid U) = \prod_{l=1}^{L} \prod_{m=1}^{M} P(T_{lm} \mid U) = \prod_{l=1}^{L} \prod_{m=1}^{M} \lambda_{lm} \exp(-\lambda_{lm} T_{lm}). \tag{11}$$

- The second alternative is to assume the attention duration $T_{lm}$ following a Gamma distribution, i.e., $T_{lm} \sim Gamma(\cdot \mid \alpha_{lm}, \beta_{lm})$, where the two parameters are regressed on the difference $\delta_m(a_{s_1^l}, a_{s_2^l})$ by taking the following forms:

$$\begin{aligned} \log \alpha_{lm} &= c_1 \delta_m(a_{s_1^l}, a_{s_2^l}) + c_2 = c_1 \left| U_m(a_{s_1^l}) - U_m(a_{s_2^l}) \right| + c_2, \\ \log \beta_{lm} &= c_3 \delta_m(a_{s_1^l}, a_{s_2^l}) + c_4 = c_3 \left| U_m(a_{s_1^l}) - U_m(a_{s_2^l}) \right| + c_4. \end{aligned} \tag{12}$$

Note that the above equations share the same parameters $c_1$, $c_2$, $c_3$, and $c_4$ with Equation (6). Then, by assuming independence among criteria, the likelihood for all attention durations in $\mathcal{T}$ conditional on the preference model $U(\cdot)$ can be written as follows:

$$P(\mathcal{T} \mid U) = \prod_{l=1}^{L} \prod_{m=1}^{M} P(T_{lm} \mid U) = \prod_{l=1}^{L} \prod_{m=1}^{M} \frac{T_{lm}^{\alpha_{lm}-1} \exp(-T_{lm}/\beta_{lm})}{\beta_{lm}^{\alpha_{lm}} \Gamma(\alpha_{lm})}. \tag{13}$$

- We can also opt for a Poisson distribution to model the attention duration $T_{lm}$ by assuming $\lceil T_{lm} \rceil \sim Poisson(\cdot \mid \lambda_{lm})$, in which $\lceil T_{lm} \rceil$ is derived by rounding the attention duration $T_{lm}$ up to the nearest integer, and parameter $\lambda_{lm}$ is regressed by taking the following form:

$$\log \lambda_{lm} = c_1 \delta_m(a_{s_1^l}, a_{s_2^l}) + c_2 = c_1 \left| U_m(a_{s_1^l}) - U_m(a_{s_2^l}) \right| + c_2, \tag{14}$$

where parameters $c_1$ and $c_2$ have the same role as in Equation (8). Then, by assuming independence among criteria, the likelihood for all attention durations $\mathcal{T}$ conditional on the preference model $U(\cdot)$ can be formulated as follows:

$$P(\mathcal{T} \mid U) = \prod_{l=1}^{L} \prod_{m=1}^{M} P(T_{lm} \mid U) = \prod_{l=1}^{L} \prod_{m=1}^{M} \frac{(\lambda_{lm})^{\lceil T_{lm} \rceil} \exp(-\lambda_{lm})}{\lceil T_{lm} \rceil!}. \tag{15}$$

We assume the above three distributions share the same regression parameters $c_1$, $c_2$, $c_3$, and $c_4$ with the distributions employed for modeling the response time in Section 2.2.2. This is reasonable because they play the same roles in measuring the DM's sensitivity to (marginal) value difference and individual heterogeneity in processing speed. Moreover, this assumption maintains the flexibility of the proposed framework in modeling different types of preference cues while avoiding the problem of over-fitting caused by incorporating too many parameters.

*2.2.4. Approximating marginal value functions with piecewise-linear functions*

The true value function model $U(\cdot)$ of the DM remains unknown a priori. Hence, we need a technique to approximate it. There are diverse types of marginal value functions (including linear, piecewise-linear, splined, and general monotone) that can be utilized for this approximation (see [55]). For simplicity, we only consider the case of piecewise-linear functions. Approximating the true value function model under such a hypothesis can be seen as a nonparametric technique because we do not need to make any assumptions on the form of the true marginal value functions [55]. Such an approximation technique has been widely used in MCDA (see, e.g., [34, 54]). We present a detailed elaboration on this technique in eAppendix A. With the approximation using piecewise-linear marginal value function, the marginal value function $U_m(\cdot)$ can be formulated as:

$$U_m(g_m(a_n)) = \boldsymbol{u}_j^\top \boldsymbol{V}_m(a_n), \tag{16}$$

where $\boldsymbol{u}_j \in \mathbb{R}^{\gamma_j}$ is the parameter associated with the marginal value function $U_m(\cdot)$, $\boldsymbol{V}_m(a_n) \in \mathbb{R}^{\gamma_j}$ is a characteristic vector on criterion $g_m$, and $\gamma_j$ is the number of subintervals on criterion $g_m$. Then, by gathering all $\boldsymbol{u}_j$ and $\boldsymbol{V}_m(a_n)$ to denote $\boldsymbol{u} = (\boldsymbol{u}_1^\top, \ldots, \boldsymbol{u}_m^\top)^\top$, and $\boldsymbol{V}(a_n) = (\boldsymbol{V}_1(a_n)^\top,$
$\ldots, \boldsymbol{V}_m(a_n)^\top)^\top$, the true value function model $U(\cdot)$ can be specified as:

$$U(a_n) = \boldsymbol{u}^\top \boldsymbol{V}(a_n), \tag{17}$$

where $\boldsymbol{V}(a_n)$ is a characteristic vector uniquely determined by the performance of alternative $a_n$ across multiple criteria, and $\boldsymbol{u}$ is the only parameter vector required to compute the comprehensive value of each alternative. Note that the parameter vector $\boldsymbol{u}$ characterizes the intrinsic nature of the value function model $U(\cdot)$. It is defined over the $(\gamma-1)$-dimensional simplex ($\gamma = \sum\limits_{m=1}^{M} \gamma_m$), which ensures the monotonicity and normalization properties of the value function model $U(\cdot)$. In essence, the DM's preference model $U(\cdot)$ is parameterized by the vector $\boldsymbol{u}$. Consequently, we can use $\boldsymbol{u}$ to signify $U(\cdot)$ in subsequent discussions. Following this representation, the value difference $\delta(a_{s_1^l}, a_{s_2^l}) = \left| U(a_{s_1^l}) - U(a_{s_2^l}) \right|$ and $\delta_m(a_{s_1^l}, a_{s_2^l}) = \left| U_m(a_{s_1^l}) - U_m(a_{s_2^l}) \right|$ can be also expressed, respectively, as:

$$\delta(a_{s_1^l}, a_{s_2^l}) = \left| U(a_{s_1^l}) - U(a_{s_2^l}) \right| = \left| \boldsymbol{u}^\top \boldsymbol{V}(a_{s_1^l}) - \boldsymbol{u}^\top \boldsymbol{V}(a_{s_2^l}) \right|, \tag{18}$$

and

$$\delta_m(a_{s_1^l}, a_{s_2^l}) = \left| U_m(a_{s_1^l}) - U_m(a_{s_2^l}) \right| = \left| \boldsymbol{u}_m^\top \boldsymbol{V}_m(a_{s_1^l}) - \boldsymbol{u}_m^\top \boldsymbol{V}_m(a_{s_2^l}) \right|. \tag{19}$$

Additionally, the prior $P(U)$ and the likelihood terms $P(\mathcal{I} \mid U)$, $P(\mathcal{R} \mid U)$, and $P(\mathcal{T} \mid U)$ are represented as $P(\boldsymbol{u})$, $P(\mathcal{I} \mid \boldsymbol{u})$, $P(\mathcal{R} \mid \boldsymbol{u})$, and $P(\mathcal{T} \mid \boldsymbol{u})$, respectively.

*2.2.5. Specifying prior distributions for model parameters*

The Bayesian inference necessitates specifying prior distributions for all the parameters involved in the assumed model. These parameters include the parameter vector $\boldsymbol{u}$ for characterizing the value function model and the regression parameter vector $\boldsymbol{c}$ ($\boldsymbol{c} = (c_1, c_2)$ for the cases of the exponential distribution and the Poisson distribution; $\boldsymbol{c} = (c_1, c_2, c_3, c_4)$ for the case of the Gamma distribution).

For the parameter vector $\boldsymbol{u}$ being defined over the $(\gamma - 1)$-dimensional simplex, we can specify a $\gamma$-dimensional Dirichlet distribution for $\boldsymbol{u}$. Such a specification has been adopted in previous MCDA literature (see, e.g., [63, 64]). Formally, the prior is given as follows:

$$P(\boldsymbol{u}) = \mathrm{Dir}(\boldsymbol{u} \mid \boldsymbol{\tau}) = \frac{1}{B(\boldsymbol{\tau})} \prod_{i=1}^{\gamma} u_i^{\tau_i - 1}, \tag{20}$$

where $u_i$ is the $i$-th entry of the vector $\boldsymbol{u}$, and $\boldsymbol{\tau} = (\tau_1, \ldots, \tau_\gamma)$ is the hyperparameter for the Dirichlet distribution such that $\tau_i > 0$ for $i = 1, \ldots, \gamma$. Also, $B(\boldsymbol{\tau}) = \frac{\prod_{i=1}^{\gamma} \Gamma(\tau_i)}{\Gamma(\Sigma_{i=1}^{\gamma} \tau_i)}$ is the normalization factor, and $\Gamma(\cdot)$ is the Gamma function (see [61]). This specification ensures any instance of the vector $\boldsymbol{u}$ generated by the prior distribution $\mathrm{Dir}(\boldsymbol{u} \mid \boldsymbol{\tau})$ conditional on the hyperparameter $\boldsymbol{\tau}$ resides in the $(\gamma-1)$-dimensional simplex, such that the monotonicity and normalization properties of the value function model $U(\cdot)$ are fulfilled. In particular, we set $\boldsymbol{\tau}$ to be a vector with all entries being equal to one. In this setting, $\mathrm{Dir}(\boldsymbol{u} \mid \boldsymbol{\tau})$ boils down to a uniform distribution defined over the $(\gamma - 1)$-dimensional simplex. Laplace's principle of insufficient reason [61] supports such a choice when there is no other rationale to favor one value function model over another in the stage of specifying a prior.

Then, for the regression parameter vector $\boldsymbol{c}$, we opt for a hierarchical prior. The reason for this choice derives from $\boldsymbol{c}$ including individual-level parameters that reflect the heterogeneity of different DMs in processing decision-related information. The advantage of specifying a hierarchical prior for $\boldsymbol{c}$ is to avoid subjective bias in inferring individual-level preferences caused by improper specification of hyperparameter values. Such a method has been widely used in Bayesian models [61] and recently adopted in MCDA literature (see, e.g., [57]). Specifically, the hierarchical prior is implemented by putting a second-stage prior on a first-stage prior. The first-stage prior specifies a multivariate Gaussian prior for the regression parameter vector $\boldsymbol{c}$ as follows:

$$P(\boldsymbol{c}) = \text{MultiNorm}(\boldsymbol{c} \mid \boldsymbol{\mu}, \boldsymbol{\Sigma}) = \frac{1}{(2\pi)^{k/2} |\boldsymbol{\Sigma}|^{1/2}} \exp\left(-\frac{1}{2}(\boldsymbol{c} - \boldsymbol{\mu})^\top \boldsymbol{\Sigma}^{-1}(\boldsymbol{c} - \boldsymbol{\mu})\right), \tag{21}$$

where $\boldsymbol{\mu}$ is the mean vector and $\boldsymbol{\Sigma}$ is the covariance matrix of the multivariate Gaussian distribution $\text{MultiNorm}(\boldsymbol{c} \mid \boldsymbol{\mu}, \boldsymbol{\Sigma})$. Here, $k$ equals 4 for the Gamma distribution, or $k$ equals 2 for the exponential and Poisson distributions. Note that $\boldsymbol{\mu}$ and $\boldsymbol{\Sigma}$ are not manually specified but are estimated through Bayesian inference. For this purpose, the second-stage prior imposes a multivariate Gaussian distribution as the prior for the mean vector $\boldsymbol{\mu}$ and an inverse Wishart distribution as the prior for the covariance matrix $\boldsymbol{\Sigma}$:

$$P(\boldsymbol{\mu}) = \text{MultiNorm}(\boldsymbol{\mu} \mid \boldsymbol{\zeta}, \boldsymbol{\Gamma}) = \frac{1}{(2\pi)^{k/2} |\boldsymbol{\Gamma}|^{1/2}} \exp\left(-\frac{1}{2}(\boldsymbol{\mu} - \boldsymbol{\zeta})^\top \boldsymbol{\Gamma}^{-1}(\boldsymbol{\mu} - \boldsymbol{\zeta})\right), \tag{22}$$

$$P(\boldsymbol{\Sigma}) = \text{InvWishart}(\boldsymbol{\Sigma} \mid \epsilon, \boldsymbol{\Psi}) = \frac{|\boldsymbol{\Psi}|^{\epsilon/2}}{2^{\epsilon k/2} \Gamma_k\left(\frac{\epsilon}{2}\right)} |\boldsymbol{\Sigma}|^{-(\epsilon+k+1)/2} \exp\left(-\frac{1}{2}\text{tr}(\boldsymbol{\Psi}\boldsymbol{\Sigma}^{-1})\right), \tag{23}$$

where $\boldsymbol{\zeta}$ is the mean vector and $\boldsymbol{\Gamma}$ is the covariance matrix of the multivariate Gaussian distribution $\text{MultiNorm}(\boldsymbol{\mu} \mid \boldsymbol{\zeta}, \boldsymbol{\Gamma})$, and $\epsilon$ is the degree of freedom and $\boldsymbol{\Psi}$ is the scale matrix of the inverse Wishart distribution $\text{InvWishart}(\boldsymbol{\Sigma} \mid \epsilon, \boldsymbol{\Psi})$, and $\Gamma_k(\cdot)$ is the $k$-dimensional multivariate Gamma function (see [61]). Such a two-stage hierarchical prior allows for flexible modeling of the uncertainty in the regression parameter vector $\boldsymbol{c}$, thus accommodating the heterogeneity of different DMs in processing decision-related information. As suggested by [62], the hyperparameters $\boldsymbol{\zeta}$, $\boldsymbol{\Gamma}$, $\epsilon$ and $\boldsymbol{\Psi}$ are set to be $\boldsymbol{0}$, $100\boldsymbol{I}$, $k+2$, and $10^{-2}\boldsymbol{I}$, respectively, where $\boldsymbol{I}$ is an identity matrix.

*2.3. Generative process and posterior inference*

An overall structure of the proposed approach is illustrated by a probabilistic graphical model shown in Figure 1. The corresponding generative process can be summarized as follows:

1. Generate value function model $\boldsymbol{u} \sim \text{Dir}(\boldsymbol{u} \mid \boldsymbol{\tau})$;
2. Generate parameters $\boldsymbol{\mu} \sim \text{MultiNorm}(\cdot \mid \boldsymbol{\zeta}, \boldsymbol{\Gamma})$, $\boldsymbol{\Sigma} \sim \text{InvWishart}(\cdot \mid \epsilon, \boldsymbol{\Psi})$;
3. Generate regression parameter vector $\boldsymbol{c} \sim \text{MultiNorm}(\boldsymbol{\mu}, \boldsymbol{\Sigma})$;
4. For each piece of preference information $q_l \in \mathcal{Q}$,
   (a) generate pairwise comparison $I_l \sim P(I_l : a_{s_1^l} \succ a_{s_2^l} \mid \boldsymbol{u})$;
   (b) derive auxiliary variable $\delta(a_{s_1^l}, a_{s_2^l}) = \left|\boldsymbol{u}^\top \boldsymbol{V}(a_{s_1^l}) - \boldsymbol{u}^\top \boldsymbol{V}(a_{s_2^l})\right|$;
   (c) generate response time $R_l$ based on Equation (5), (7), or (9);
   (d) For each criterion $g_m \in \mathcal{G}$:
      i. derive auxiliary variable $\delta_m(a_{s_1^l}, a_{s_2^l}) = \left|\boldsymbol{u}_m^\top \boldsymbol{V}_m(a_{s_1^l}) - \boldsymbol{u}_m^\top \boldsymbol{V}_m(a_{s_2^l})\right|$;
      ii. generate attention duration $T_{lm}$ based on Equation (11), (13), or (15).

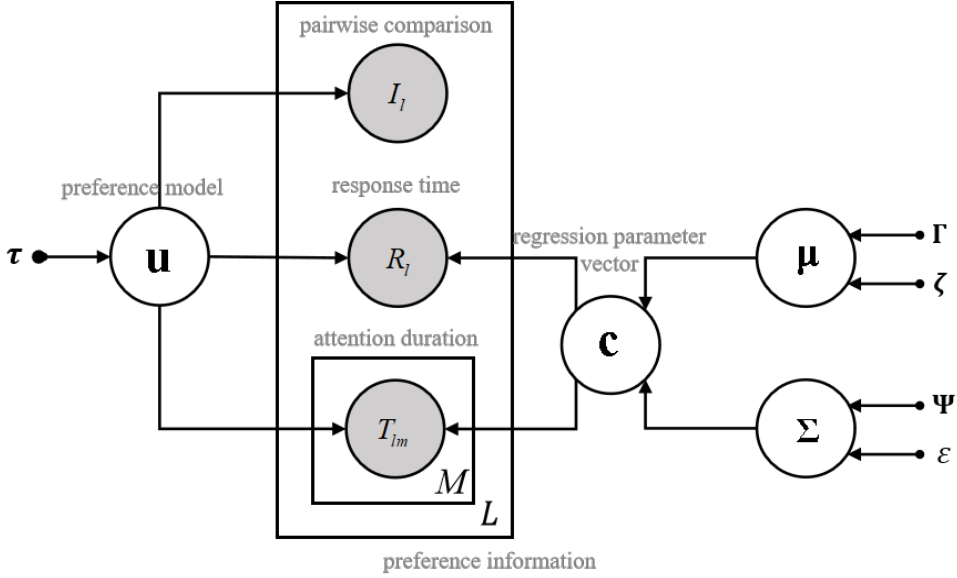

Figure 1: Directed acyclic graph of our model.

We adopt a full Bayesian approach to derive the posterior distribution of the proposed model. Let $\theta = \{\zeta, \Gamma, \epsilon, \Psi, \tau\}$ represent all hyperparameters. The expression in Equation (2) can then be reformulated as follows:

$$
\begin{aligned}
P(\boldsymbol{u}, &\boldsymbol{c}, \boldsymbol{\mu}, \boldsymbol{\Sigma} \mid \mathcal{I}, \mathcal{R}, \mathcal{T}, \theta) \\
=& P(\boldsymbol{u} \mid \boldsymbol{\tau}) P(\boldsymbol{\mu} \mid \boldsymbol{\zeta}, \boldsymbol{\Gamma}) P(\boldsymbol{\Sigma} \mid \epsilon, \boldsymbol{\Psi}) P(\boldsymbol{c} \mid \boldsymbol{\mu}, \boldsymbol{\Sigma}) \\
& \times \prod_{l=1}^{L} P(I_l \mid \boldsymbol{u}) P(R_l \mid \boldsymbol{u}, \boldsymbol{c}) \times \prod_{l=1}^{L} \prod_{m=1}^{M} P(T_{lm} \mid \boldsymbol{u}, \boldsymbol{c}) \\
& / P(\mathcal{I}, \mathcal{R}, \mathcal{T} \mid \theta) \\
=& P(\boldsymbol{u} \mid \boldsymbol{\tau}) P(\boldsymbol{\mu} \mid \boldsymbol{\zeta}, \boldsymbol{\Gamma}) P(\boldsymbol{\Sigma} \mid \epsilon, \boldsymbol{\Psi}) P(\boldsymbol{c} \mid \boldsymbol{\mu}, \boldsymbol{\Sigma}) \\
& \times \prod_{l=1}^{L} P(I_l \mid \boldsymbol{u}) P(R_l \mid \boldsymbol{u}, \boldsymbol{c}) \times \prod_{l=1}^{L} \prod_{m=1}^{M} P(T_{lm} \mid \boldsymbol{u}, \boldsymbol{c}) \\
& / \iiiint P(\mathcal{I}, \mathcal{R}, \mathcal{T}, \boldsymbol{u}, \boldsymbol{c}, \boldsymbol{\mu}, \boldsymbol{\Sigma} \mid \theta) \, \mathrm{d}\boldsymbol{u} \, \mathrm{d}\boldsymbol{c} \, \mathrm{d}\boldsymbol{\mu} \, \mathrm{d}\boldsymbol{\Sigma}.
\end{aligned}
\tag{24}
$$

In this framework, the posterior distributions $P(R_l \mid \boldsymbol{u}, \boldsymbol{c})$ and $P(T_{lm} \mid \boldsymbol{u}, \boldsymbol{c})$ are adapted based on the specified distributions (exponential, Gamma, or Poisson) for response times and attention durations. Specifically, Equations (5) and (11) are used for the case of exponential distribution, Equations (7) and (13) for the case of Gamma distribution, and Equations (9) and (15) for the case of Poisson distribution.

Due to the intricate dependency structure inherent within the normalization term $P(\mathcal{I}, \mathcal{R}, \mathcal{T} \mid \theta)$ in Equation (24), it is not feasible to compute the posterior distribution analytically. This necessitates the use of approximate inference methods. Traditional approaches, such as Markov Chain Monte Carlo (MCMC), are often employed to summarize the posterior distribution by iterative sampling from full conditional distributions and making inferences based on the aggregated random samples [62]. However, the non-conjugate priors employed in our model introduce significant challenges for implementing standard MCMC algorithms, particularly in achieving a stationary distribution. To overcome these challenges, we utilize Hamiltonian Monte Carlo (HMC), a modern variant of MCMC [39]. HMC exploits first-order gradient information to accelerate convergence to high-dimensional target distributions, offering a significant improvement in efficiency compared to traditional MCMC methods. This algorithm is implemented within NumPyro, a probabilis-tic programming library that provides a flexible and efficient framework for Bayesian modeling. NumPyro supports automatic differentiation for gradient-based optimization and enables scalable inference through parallel computation, making it well-suited for handling the complexity of our model. We only need to specify the conditional dependencies among parameters, and NumPyro automates the inference process using the HMC algorithm.

*2.4. Outline of the process*

The approach follows a structured six-step process, outlined in Figure 2. Each step is detailed below:

*Step 1: Problem definition.* Define the decision problem by specifying the set of alternatives $\mathcal{A}$, the set of criteria $\mathcal{G}$, and the performance of alternatives against each criterion.

*Step 2: Preference elicitation procedure.* The DM is required to provide pairwise comparisons $\mathcal{I} = \{I_1, \ldots, I_l, \ldots, I_L\}$, alongside supplementary information, including response times $\mathcal{R} = \{R_1, \ldots, R_l, \ldots, R_L\}$ and attention durations $\mathcal{T} = \{T_{11}, \ldots, T_{lm}, \ldots, T_{LM}\}$.

*Step 3: Build posterior distribution.* Apply the Bayes' rule to construct a posterior distribution $P(\boldsymbol{u}, \boldsymbol{c}, \boldsymbol{\mu}, \boldsymbol{\Sigma} \mid \mathcal{I}, \mathcal{R}, \mathcal{T}, \theta)$.

*Step 4: Parameter inference.* Apply the HMC algorithm to collect samples of $\boldsymbol{u}, \boldsymbol{c}, \boldsymbol{\mu}, \boldsymbol{\Sigma}$ from the posterior distribution $P(\boldsymbol{u}, \boldsymbol{c}, \boldsymbol{\mu}, \boldsymbol{\Sigma} \mid \mathcal{I}, \mathcal{R}, \mathcal{T}, \theta)$.

*Step 5: Convergence analysis.* Analyze whether the Markov chain has converged to its stationary distribution. If not, reinitialize the chain for resampling.

*Step 6: Result analysis.* Analyze the collected samples of $\boldsymbol{u}, \boldsymbol{c}, \boldsymbol{\mu}, \boldsymbol{\Sigma}$ to compute stochastic acceptabilities summarizing the support for specific ranking and choice results [51]. The primary outcomes we consider are Pairwise Winning Indices that reflect the shares of samples confirming the truth of preference relation for each pair and Rank Acceptability Indices estimating the size of preference model subspace in which an alternative is assigned some rank [45]. Additionally, we summarize the posterior distribution of the inferred model.

## 3. Review of experimental studies with real-world subjects in MCDA

Experiments in behavioral economics with real-world subjects have profoundly reshaped our understanding of decision-making. For example, the seminal works by Kahneman and Tversky [44, 73] demonstrated that individuals often deviate from the rational actor model through cognitive biases and framing effects. Their pioneering studies on prospect theory, heuristics, and biases revealed that evaluating risk and uncertainty is far more complex than traditional economic theories suggest. In parallel, Slovic's research has illuminated the influence of affect and perception in risk assessment, showing that subjective experiences often drive decision outcomes [68]. These foundational insights have also laid the groundwork for experimental studies involving real-world subjects in MCDA.

Such experiments bridge the gap between theoretical models and practical applications. Unlike purely mathematical analyses or simulations, they allow researchers to assess how DMs interact with MCDA methods in authentic contexts. In what follows, we discuss a representative subset of such studies while grouping them by their primary aims. Their characteristics, including involved subjects, tackled problems, and employed methods, are summarized in Table 1.

Some studies have focused on whether specific MCDA methods can effectively support real-world decision-making. For example, Ishizaka et al. [41], Korhonen and Topdagi [48], and Huizingh and Vrolijk [40] evaluated the performance of the Analytic Hierarchy Process (AHP) in replicating DMs' inherent preferences. The experiments demonstrated that the method could provide recommendations that closely align with the directly revealed choices or implicit preferences (even when the ratio scale assumption is violated) or outperform some naive procedures that do not account for multiple criteria characteristics of alternatives.

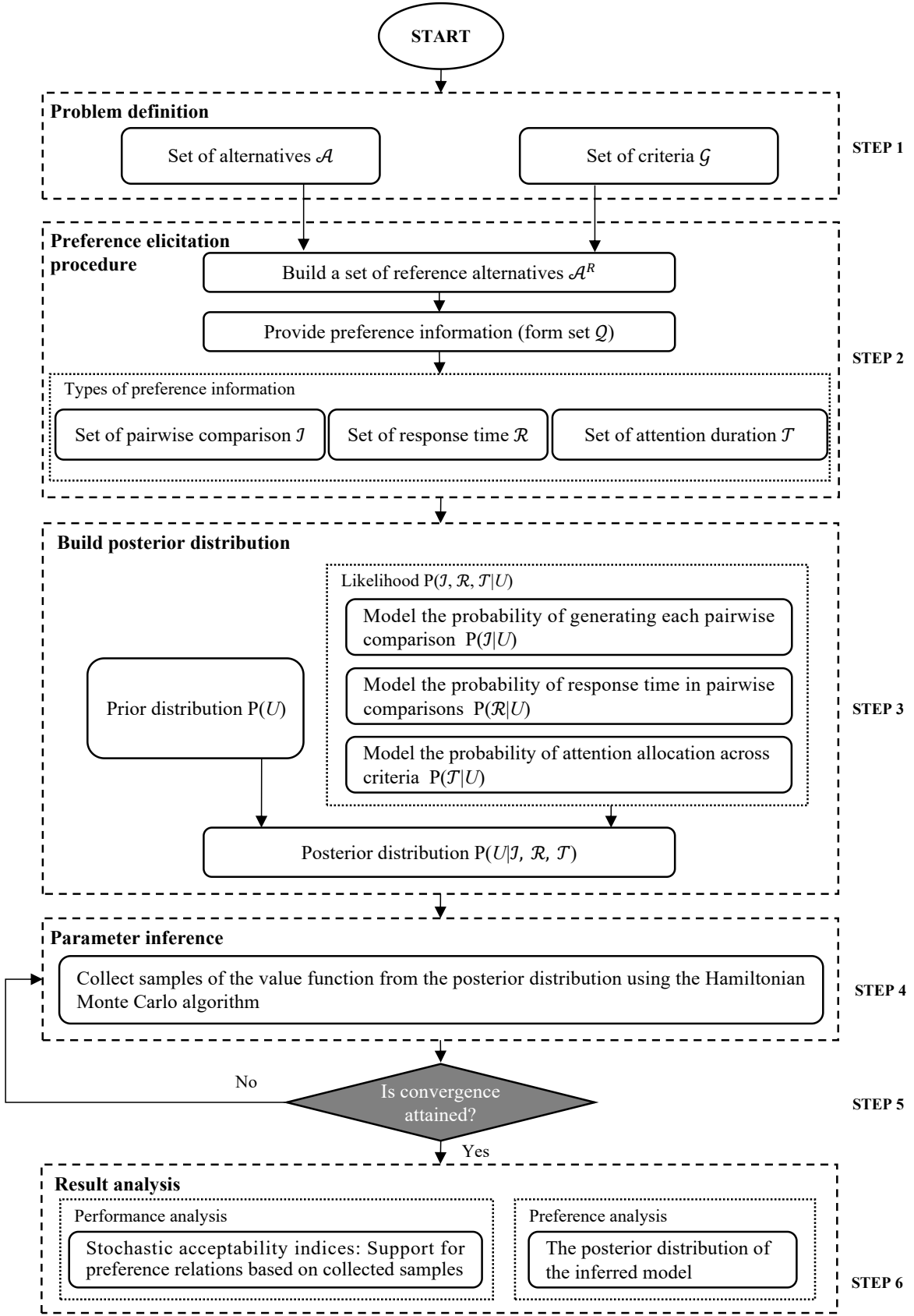

Figure 2: Framework of the proposed approach.

Table 1: Overview of experiments with real-world subjects using various MCDA methods.

| Study | Real-world subjects | Decision problem | Multiple criteria methods | Aim |
| --- | --- | --- | --- | --- |
| Weber et al. [75] (1988) | 256 graduate students from business school in Aachen | Evaluation of a future job | Multi-attribute utility theory using value trees | Examine the effect of splitting objectives on weights |
| Keeney et al. [47] (1990) | 9 engineers and 14 social science teachers | Assessing long-term energy policies | Public value forum combining focus groups and direct multi-attribute value elicitation techniques | Examine the feasibility of eliciting lay values and comparing them with expert assessments |
| Borcherding et al. [12] (1991) | 200 students from the University of Southern California | Evaluation of nuclear waste repository sites | Ratio, swing weighting, tradeoff, and pricing out methods | Compare weighting methods on internal consistency, convergent and external validity |
| Salminen and Wallenius [66] (1993) | 48 management students from the University of Jyvaskyla | Pairwise comparisons of two-bedroom condominiums | Prospect theory and traditional value model | Compare the methods in terms of explaining observed preferences |
| Hobbs and Meier [38] (1994) | 12 company staff (Seattle City Light) with different roles | Ranking hypothetical portfolios of resources | Direct weight assessment, tradeoff weight assessment, additive value functions, and goal programming | Compare MCDA methods in terms of ease of use, appropriateness, sensitivity, matching preferences, yielding the same weights and rankings |
| Huizingh and Vrolijk [40] (1997) | 180 students from the University of Groningen | Selection of a room to rent | AHP variant using absolute intensity ratings | Test the applicability of AHP in repetitive decision-making and its superiority over a naive model |
| Brugha [15] (2000) | 10 students from the University College Dublin | Choosing a career or buying a car | AHP (relative measurement) and SMART (scoring with intervals) | Evaluate scoring methods and propose a DM-centred approach for method selection |
| Korhonen and Topdagi [48] (2003) | 8 high-school students from Finland | Evaluation of vegan vs. non-vegan meals | AHP applied in a setting with mixed utility/disutility comparisons | Assess AHP's performance when the ratio scale assumption is violated |
| Brugha [16] (2004) | 53 students from the University College Dublin | Career choice with 2 to 6 alternatives | DISCUSS, DISCRIM, and SMART | Explore the appropriateness of methods at various decision stages |
| Hämäläinen and Alaja [37] (2008) | 30 students, local citizens, experts, and summer residents | Regulation of a lake–river system | Attribute weighting with different value tree structures | Test for splitting bias and its possible reduction through instruction/training |
| Linares [53] (2009) | 18 graduate students from Spanish Polytechnics | Comparing five compact cars (global and aesthetics) | Pairwise comparisons using AHP | Investigate whether removing intransitivities improves the representation of DMs' true preferences |
| Ishizaka et al. [41] (2011) | 21 undergraduates from the University of Exeter | Choosing among five boxes of chocolates | AHP with pairwise comparisons and multiple ranking evaluations | Assess AHP as a choice method and its consistency in replicating preferences |
| Korhonen et al. [50] (2012) | 144 sophomores at the Helsinki School of Economics | Binary choices based of study program implementations | Pairwise comparisons in a bi-criteria problem | Examine consistency with a linear value function |
| Beccacece et al. [8] (2015) | 65 students from the Bocconi University | Evaluation of mobile phone packages | Elicitation of multi-attribute value functions via pairwise comparisons | Assess interactions and monotonicity in elicited value functions |
| Deparis et al. [27] (2015) | 29 graduate students from École Centrale Paris | Evaluation of apartments to rent | Bi-matching (forward and reverse) for preference elicitation | Investigate the effect of multi-criteria conflict on matching judgments |
| Ciomek et al. [21] (2017) | 101 students from Poznań University of Technology | Ranking 10 mobile phone packages | Strategies for prioritizing pairwise elicitation questions in UTA methods | Compare robustness of recommended rankings and interaction time |
| Ishizaka and Siraj [42] (2018) | 146 students and university staff in the United Kingdom | Comparing five coffee shops | AHP, SMART, and MACBETH | Evaluate the methods' usefulness and the effect of recommendations on ranking changes |
| Gehrlein et al. [32] (2023) | 115 real nominators from the Polkadot ecosystem | Selection of active validators in blockchain environment | Active learning strategies for UTA methods and non-supported selection | Compare interaction time, cognitive effort, and efficacy |
| This paper | 30 students from Xi'an Jiaotong University | Ranking 10 mobile phone contracts | BOR variants fed with pairwise comparisons and behavioral cues | Compare consistency in reconstructing DM's preferences and uncover the role of behavioral cues |

Another critical area of research involves comparative evaluations of MCDA methods and tests of preference elicitation techniques. For instance, Borcherding et al. [12] and Hobbs and Meier [38] compared various weighting methods based on internal consistency, convergent validity, and external validity. Furthermore, Ishizaka and Siraj [42] compared AHP, SMART, and MACBETH, assessing how different approaches influenced DMs' rankings. More recently, studies by Ciomek et al. [21] and Gehrlein et al. [32] focused on active learning strategies for preference elicitation, exploring ways to minimize cognitive effort and interaction time. In turn, Deparis et al. [27] explored how forward and reverse matching methods could be used for preference elicitation in scenarios with different levels of multiple criteria conflict or trade-off size. Finally, Brugha [15] and Brugha [16] tested methods, such as AHP, SMART, DISCUSS, and DISCRIM, emphasizing how they may be suitable at different stages of decision processes.

Some research has focused on the overall capability of MCDA techniques to capture complex human judgments accurately. In this context, Korhonen et al. [50] examined whether linear value functions can reliably reproduce DM's pairwise comparisons in bi-criteria. A related experiment concerning the interpretation of criteria weights is presented in [49]. In turn, Beccacece et al. [8] explored how multi-attribute value functions could be elicited through pairwise comparisons to assess monotonicity and interaction effects in preferences. Another strand of research is instrumental in bridging behavioral theory with practical MCDA applications, enriching our understanding of how psychological, cognitive, and behavioral factors shape decision-making processes. In particular, Salminen and Wallenius [66] tested prospect theory within an MCDA framework, comparing its predictive power against traditional value models. Similarly, Linares [53] explored the impact of intransitivities on decision outcomes, thereby questioning whether enforcing consistency leads to more valid representations of DMs' preferences.

A related research objective is identifying and mitigating biases inherent in the elicitation process, as biases can significantly distort decision outcomes. Weber et al. [75] explored how splitting objectives in a value tree structure affects weight assignments and whether it leads to overweighting bias. Similarly, Hämäläinen and Alaja [37] examined the effects of value tree structuring on DMs' weight assessments, testing whether training and instructions could reduce biases. Keeney et al. [47] took a different approach by investigating the feasibility of eliciting public values for energy policy, emphasizing how laypeople's inputs compared with expert assessments. Some other bias detection and debiasing experiments are listed in [60]. Still, this review paper concludes that testing best practices to reduce cognitive and motivational biases in controlled experiments is of high priority.

We present an original experiment with real-world subjects with multiple aims. First, we compare different methods within ordinal regression and Bayesian preference inference frameworks regarding their ability to reconstruct the DM's comprehensive preferences and provide robust recommendations. Also, we assess the methods' capabilities for representing DMs' preferences and verify if incorporating behavioral indicators improves preference modeling and inference. Finally, we test behavioral theories in MCDA contexts by checking if response time and attention allocation reflect cognitive mechanisms and influence decision-making.

## 4. Experiment involving real subjects

*4.1. Experimental design*

*4.1.1. Experimental task*

In the experiment, participants compared 10 real mobile phone contracts offered by three Chinese network operators. These contracts were described using two gain-type criteria and four cost-type criteria: $g_1$ (domestic calls), the free call time included for domestic usage; $g_2$ (domestic data), the amount of free data included in the contract; $g_3$ (overage call fee), the cost per additional minute of domestic calls beyond the contract's included minutes; $g_4$ (overage data fee), the cost per additional GB of data used beyond the contract's allowance; $g_5$ (monthly fee), the basic fee paid to the operator each month; $g_6$ (initial deposit), the upfront fee required when selecting the contract. The corresponding performance matrix for this decision-making task is presented in Table 2. Note that, in general, the feasibility of answering indirect questions and the underlying cognitive effort depend on the number of criteria involved [28]. Yet, pairwise comparisons with six criteria – as used in our study – are deemed appropriate and feasible for DMs to answer accurately [65].

Table 2: Performances of 10 real phone contracts in terms of six criteria.

|  | Domestic calls (min.) $g_1$(gain) | Domestic data (GB) $g_2$(gain) | Overage call fee (RMB/min.) $g_3$(cost) | Overage data fee (RMB/GB) $g_4$(cost) | Monthly fee $g_5$(cost) | Initial deposit $g_6$(cost) |
|---|---|---|---|---|---|---|
| $a_1$ | 150 | 15 | 0.19 | 3 | 79 | 100 |
| $a_2$ | 50 | 3 | 0.25 | 10 | 29 | 200 |
| $a_3$ | 700 | 40 | 0.13 | 3 | 169 | 0 |
| $a_4$ | 1000 | 60 | 0.1 | 3 | 199 | 0 |
| $a_5$ | 500 | 30 | 0.16 | 3 | 129 | 0 |
| $a_6$ | 50 | 5 | 0.23 | 10 | 39 | 150 |
| $a_7$ | 100 | 10 | 0.21 | 5 | 59 | 100 |
| $a_8$ | 0 | 15 | 0.23 | 5 | 39 | 200 |
| $a_9$ | 200 | 20 | 0.18 | 5 | 99 | 200 |
| $a_{10}$ | 300 | 20 | 0.15 | 3 | 119 | 0 |

We paired the aforementioned 10 real-world phone contracts and excluded pairwise comparisons with strict dominance relation. From the remaining pairs, 30 were randomly selected to collect participants' preference data. Participants completed the task using a dedicated desktop application in separate laboratory sessions conducted at different times over two weeks. Specifically, they answered 30 questions by choosing between two alternatives based on their performance across various criteria (see Figures 3a–3c).

*4.1.2. Experimental procedure*

Throughout the experiment, participants' eye movements were recorded using a *Tobii T120 eye tracker* with a sampling rate of 120 Hz, integrated into a 17-inch monitor. Before the experiment began, participants were required to sign an informed consent form outlining the purpose of the experiment, data usage policies, and privacy protection measures. After signing, they were guided to a designated computer station where detailed operational instructions were provided. First, the system measured each participant's interpupillary distance and dominant eye, followed by calibrating the eye tracker and adjusting their sitting posture to ensure accurate data collection. Participants were instructed to minimize head movements during the entire experiment to maintain data accuracy. Subsequently, the screen displayed task instructions and an example interface, enabling participants to familiarize themselves with the experiment's layout and procedure. To confirm their understanding, participants were required to answer a comprehension test after reading the instructions. Only those who passed the test could proceed to the formal experiment (see Figures 3d–3f).

During the formal experimental tasks, participants made decisions by moving the mouse cursor to an option located in either the bottom-left or bottom-right corner of the screen and clicking it. The system not only logged the selected option for each click but also collected the following data via the *Tobii eye tracker*: attention duration for each criterion, response time for each task, and the $(x, y)$ coordinates of the mouse cursor on the interface. After completing the choice tasks, participants were asked to fill out a questionnaire regarding their average monthly call duration, data usage, and demographic information. Upon completion of the questionnaire, participants received a monetary reward as compensation for their time and participation in the experiment.

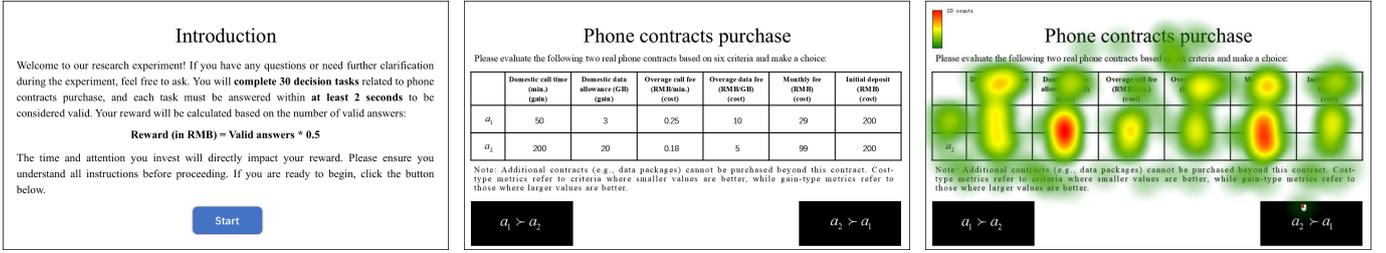

| (a) Welcome screen | (b) Pairwise elicitation question | (c) Attention duration data |

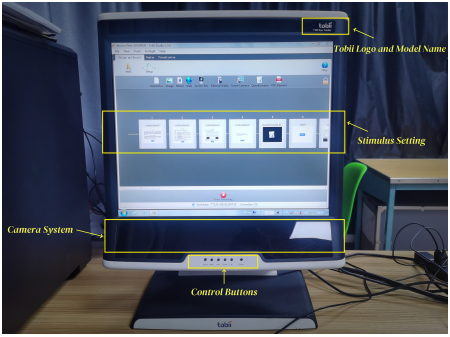
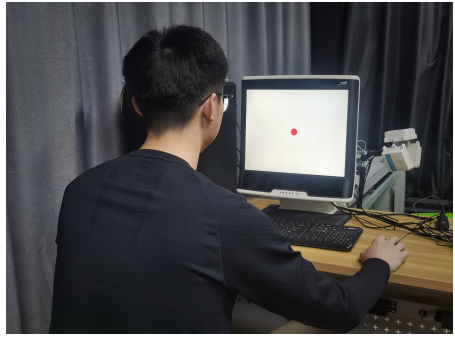
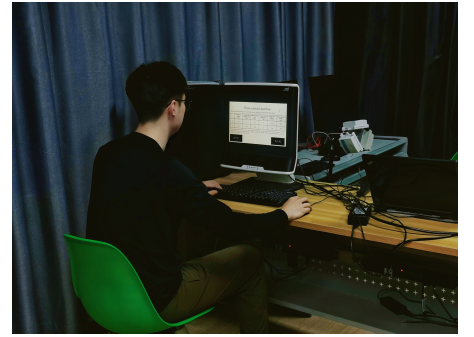

| (d) Eye movement recording | (e) Calibration for eye tracking | (f) Experimental site |

Figure 3: Interfaces of the dedicated desktop application.

*Notes.* (3a) Participants clicked the "start" button at the screen center to enter the decision stage. (3b) The decision stage displayed a performance matrix with two alternatives and two options: $a_1 \succ a_2$ (bottom-left) and $a_2 \succ a_1$ (bottom-right). (3c) Participants selected their preferred option, and the program recorded the time spent on each criterion. Heatmaps generated from eye-tracking data visualize attention durations across criteria. (3d) Eye movements were recorded using a *Tobii T120 infrared eye tracker* integrated into a 17-inch monitor with a 120 Hz sampling rate. (3e) Pupillary distance and dominant eye were measured and calibrated. (3f) Schematic of a participant performing the experiment. *Note: Text translated from Mandarin and enlarged for display.*

### 4.1.3. Participants

A total of 30 Chinese students from Xi'an Jiaotong University participated in this experiment. To be eligible, participants were required to meet the following criteria: (a) no specific eye problems (e.g., lazy eye, nystagmus, etc.), (b) no prior involvement in similar experiments within the past six months, and (c) familiarity with or prior experience in purchasing mobile phone contracts. The final sample had a mean age of 25.87 years (standard deviation (SD) = 2.20; min = 22; max = 29), with 15 participants (50%) being female.

### 4.2. Experiment results

#### 4.2.1. Statistical analysis of behavioral results

In this section, we analyze the behavioral data collected from 30 participants, including the average response time for each pairwise comparison task and the duration of attention allocated to different criteria. These analyses offer insights into participants' behavioral patterns observed during the experiment.

Table 3 summarizes the descriptive statistics for attention durations across six criteria ($g_1$ to $g_6$) and response times in pairwise comparison tasks. The longest mean attention duration is observed for $g_5$ (4.40 seconds), while the shortest is for $g_6$ (1.44 seconds), reflecting the decreasing attention participants paid to less critical criteria. In contrast, response time shows substantial variability (mean = 28.60 seconds, SD = 36.71), indicating notable differences in cognitive effort across decision-making tasks. The maximum response time (110 seconds) suggests that some tasks involved greater complexity or uncertainty, requiring extended deliberation.

Table 3: Statistical analysis of behavioral data.

| Variable | Criterion | Mean | Standard Deviation | Maximum | Minimum |
|---|---|---|---|---|---|
| Response Time | - | 28.60 | 36.71 | 110.00 | 1.01 |
| Attention Duration | $g_1$ | 2.37 | 1.57 | 20.16 | 0.43 |
| | $g_2$ | 3.54 | 2.15 | 22.40 | 0.50 |
| | $g_3$ | 1.61 | 1.17 | 18.97 | 0.32 |
| | $g_4$ | 2.11 | 1.32 | 25.55 | 0.49 |
| | $g_5$ | 4.40 | 2.28 | 27.64 | 0.35 |
| | $g_6$ | 1.44 | 0.85 | 15.67 | 0.37 |

Figure 4a presents the distribution of attention durations across six decision criteria ($g_1$ to $g_6$), highlighting significant differences between the criteria. The left panel shows the average attention duration for 30 participants. Notably, $g_2$ and $g_5$ are associated with the longest average attention durations and the widest distributions, indicating more significant importance in decision-making. In contrast, the other criteria are associated with shorter attention durations, reflecting their lower priority. The right panel shows individual attention allocation across 30 tasks, revealing greater variability, especially for $g_4$ and $g_5$, which exhibit larger interquartile ranges. Conversely, the distributions for $g_3$ and $g_6$ are more concentrated, suggesting more consistent attention allocation for lower-priority criteria.

Similarly, Figure 4b illustrates the response time distribution for the pairwise comparison tasks. The left panel shows considerable variability in response times, with a median of approximately 20 seconds and a right-skewed distribution, indicating outliers with longer response times. The right panel breaks down response times by individual tasks (pairwise comparison indices from $I_1$ to $I_{30}$). Notably, specific tasks such as $I_1, I_{12}$, and $I_{25}$, where many participants took longer to respond, exhibit wider interquartile ranges, suggesting increased cognitive effort. In contrast, tasks such as $I_9, I_{26}$, and $I_{30}$ show more compact distributions, indicating faster and more consistent decision-making. Additionally, tasks such as $I_7, I_{19}$, and $I_{28}$ show prolonged response times for some participants, which may indicate that these comparisons involve highly similar alternatives, thereby necessitating deeper cognitive processing. The observed variability in response times underscores the complexity involved in pairwise comparisons between similar alternatives.

*4.2.2. Comparative analysis*

In this section, we conduct a comparative analysis of the results produced by our nine model variants against those generated by the Bayesian ordinal regression (BOR) model proposed by [63], as presented in Table 4. BOR is selected as the benchmark due to its alignment with the Bayesian preference modeling framework and its status as a simplified version of our approach, focusing solely on preference information from pairwise comparisons. To facilitate discussion in this section, we refer to the proposed approach as Behaviorally Augmented Bayesian Ordinal Regression (BABOR) since it incorporates multiple behavioral cues to enhance preference inference.

The dataset includes preference data from 30 participants. For each participant, their data was initially split into training (80%) and test (20%) sets. The training set was further divided into training and validation subsets. This process was repeated 20 times to ensure robustness. The training subset was used for model learning, the validation subset was employed to determine the parameter $\gamma_m$, and the test set evaluated model performance across various metrics, enabling a comprehensive comparative analysis under consistent data partitioning.

To evaluate the performance of BOR and the nine BABOR variants (I-1, I-2, I-3, II-1, II-2, II-3, III-1, III-2, and III-3), which incorporate different forms of response time and attention duration, we employ two key metrics: *Average Support of the Inferred Pairwise Winning Indices on True Pairwise Comparisons between Alternatives* (ASP) and *Accuracy Rate on Test Set* (ART). The ASP metric, as used in [63, 72],

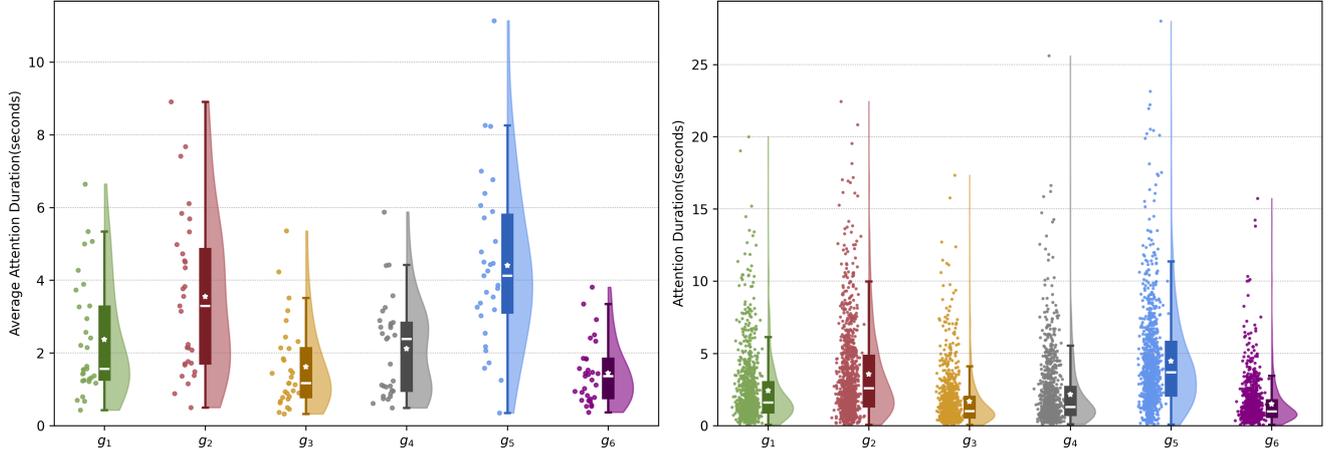

(a) Attention duration of criteria in decision-making tasks.

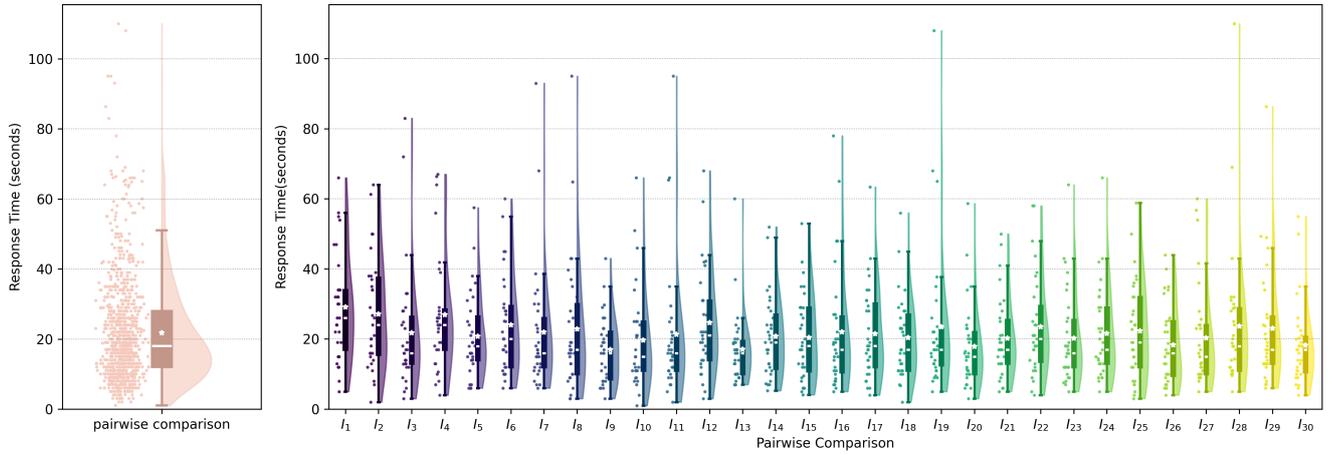

(b) Response time distribution of each pairwise comparison.

Figure 4: Statistical analysis of behavioral data.

*Notes.* Figure 4a illustrates the average attention duration across different criteria for 30 participants (left panel) and the attention duration distribution for the same participants across 30 decision-making tasks (right panel). Figure 4b presents the average response time distribution for 30 participants in each pairwise comparison task (left panel) and the detailed response time distribution across 30 pairwise comparisons (right panel). Scatter plots show individual data points, box plots summarize medians, interquartile ranges, and outliers, while distribution curves, generated using Kernel Density Estimation, provide a continuous representation of the data distribution by smoothing the data to estimate the probability density function.

Table 4: Overview of ten comparison models with types of preference information and assumptions.

| Model Name | Type of Preference Information | | | Distribution Assumption | | |
|---|---|---|---|---|---|---|
| | Pairwise Comparison | Response Time | Attention Duration | Gamma | exponential | Poisson |
| BOR | ✓ | | | | | |
| BABOR I-1 | ✓ | ✓ | ✓ | ✓ | | |
| BABOR I-2 | ✓ | ✓ | ✓ | | ✓ | |
| BABOR I-3 | ✓ | ✓ | ✓ | | | ✓ |
| BABOR II-1 | ✓ | ✓ | | ✓ | | |
| BABOR II-2 | ✓ | ✓ | | | ✓ | |
| BABOR II-3 | ✓ | ✓ | | | | ✓ |
| BABOR III-1 | ✓ | | ✓ | ✓ | | |
| BABOR III-2 | ✓ | | ✓ | | ✓ | |
| BABOR III-3 | ✓ | | ✓ | | | ✓ |

*Notes.* ✓indicates the inclusion of the corresponding type of preference information or assumption in the model.

measures the average level of agreement between inferred and true pairwise preference across all alternatives. Meanwhile, ART, as defined in [64], assesses the accuracy of inferred pairwise comparisons relative to the true preference. Comprehensive definitions and explanations of these metrics are provided in eAppendix B.

We first report the mean and standard deviation of ASP and ART for different methods in Table 5. With the exception of BABOR II-3, all proposed variants outperform the BOR model. Notably, BABOR I-1, I-2, and I-3, which incorporate both response time and attention to the criterion, consistently achieve ASP and ART values exceeding 0.8. This means that, on average, about 80% of inferred value functions confirm the true preference relation for pairs in the test set (ASP), and for about 80% of alternative pairs, more than half of the inferred models confirm the true relation (ART). Compared to BOR, BABOR I-2, the best-performing variant, improves ASP by 11.1% and ART by 11.2%, demonstrating that incorporating more comprehensive preference information leads to better alignment with the DMs' true preference. On average, compared to BOR, BABOR variants improve ASP by 5.8% and ART by 5.77%, indicating that considering additional behavioral cues provides an advantage over using only pairwise comparison information. Additionally, the standard deviations in Table 5 show that BABOR I-1, I-2, and I-3 exhibit more stable performance, with lower variability across random splits, further supporting the robustness of these methods. Still, the worst-performing BABOR variants attain average metrics' values slightly worse (II-3) or only marginally better (III-3) than BOR. These variants incorporate only one type of behavioral cue and the Poisson distribution.

Table 5: Mean and standard deviation (in round brackets) of ASP and ART across different methods.

| Metric | BOR | BABOR I-1 | BABOR I-2 | BABOR I-3 | BABOR II-1 | BABOR II-2 | BABOR II-3 | BABOR III-1 | BABOR III-2 | BABOR III-3 |
|---|---|---|---|---|---|---|---|---|---|---|
| ASP | 0.690(0.202) | 0.796(0.104) | **0.801(0.104)** | 0.800(0.103) | 0.705(0.112) | 0.788(0.102) | 0.653(0.082) | 0.720(0.153) | 0.767(0.121) | 0.702(0.106) |
| ART | 0.706(0.211) | 0.802(0.110) | **0.818(0.108)** | 0.812(0.101) | 0.715(0.121) | 0.804(0.110) | 0.695(0.092) | 0.726(0.178) | 0.779(0.116) | 0.714(0.141) |

Table 6: Results ($p$-values) of the Wilcoxon signed-rank test for the significance of differences in the average values of ASP and ART between different methods.

| Metric | Method | BOR | BABOR I-1 | BABOR I-2 | BABOR I-3 | BABOR II-1 | BABOR II-2 | BABOR II-3 | BABOR III-1 | BABOR III-2 | BABOR III-3 |
|---|---|---|---|---|---|---|---|---|---|---|---|
| ASP | BOR | - | 0.007** | 0.003** | 0.007** | 0.630 | 0.015* | 0.052 | 0.366 | 0.238 | 0.848 |
| | BABOR I-1 | 0.993 | - | 0.018* | 0.445 | 1.000 | 0.993 | 1.000 | 0.998 | 0.994 | 1.000 |
| | BABOR I-2 | 0.997 | 0.983 | - | 0.784 | 1.000 | 0.999 | 1.000 | 0.999 | 0.999 | 1.000 |
| | BABOR I-3 | 0.993 | 0.562 | 0.221 | - | 1.000 | 0.997 | 1.000 | 0.999 | 0.998 | 1.000 |
| | BABOR II-1 | 0.370 | 0.000*** | 0.000*** | 0.000*** | - | 0.000*** | 0.999 | 0.180 | 0.001** | 0.613 |
| | BABOR II-2 | 0.985 | 0.008** | 0.001** | 0.003** | 1.000 | - | 1.000 | 0.996 | 0.978 | 1.000 |
| | BABOR II-3 | 0.948 | 0.000*** | 0.000*** | 0.000*** | 0.002** | 0.000*** | - | 0.000*** | 0.000*** | 0.001** |
| | BABOR III-1 | 0.634 | 0.002** | 0.001** | 0.001** | 0.825 | 0.004** | 1.000 | - | 0.080 | 0.978 |
| | BABOR III-2 | 0.762 | 0.006** | 0.001** | 0.002** | 0.999 | 0.023* | 1.000 | 0.923 | - | 1.000 |
| | BABOR III-3 | 0.152 | 0.000*** | 0.000*** | 0.000*** | 0.395 | 0.000*** | 0.999 | 0.023* | 0.000*** | - |
| ART | BOR | - | 0.011* | 0.007** | 0.009** | 0.721 | 0.015* | 0.875 | 0.489 | 0.033* | 0.748 |
| | BABOR I-1 | 0.989 | - | 0.025* | 0.178 | 1.000 | 0.806 | 1.000 | 0.992 | 0.996 | 1.000 |
| | BABOR I-2 | 0.993 | 0.975 | - | 0.894 | 1.000 | 0.993 | 1.000 | 0.998 | 1.000 | 1.000 |
| | BABOR I-3 | 0.991 | 0.822 | 0.106 | - | 1.000 | 0.902 | 1.000 | 0.996 | 1.000 | 1.000 |
| | BABOR II-1 | 0.279 | 0.000*** | 0.000*** | 0.000*** | - | 0.000*** | 0.909 | 0.250 | 0.001** | 0.358 |
| | BABOR II-2 | 0.985 | 0.194 | 0.007** | 0.098 | 1.000 | - | 1.000 | 0.994 | 0.997 | 1.000 |
| | BABOR II-3 | 0.125 | 0.000*** | 0.000*** | 0.000*** | 0.091 | 0.000*** | - | 0.028* | 0.000*** | 0.109 |
| | BABOR III-1 | 0.511 | 0.008** | 0.002** | 0.004** | 0.762 | 0.006** | 0.972 | - | 0.135 | 0.980 |
| | BABOR III-2 | 0.967 | 0.004** | 0.000*** | 0.000*** | 0.999 | 0.003** | 1.000 | 0.865 | - | 0.998 |
| | BABOR III-3 | 0.252 | 0.000*** | 0.000*** | 0.000*** | 0.642 | 0.000*** | 0.891 | 0.020* | 0.002** | - |

*Notes.* *, ** and *** denote significance levels based on the Wilcoxon signed-rank test, indicating that the column model significantly outperforms the row model at $p < 0.05$, $p < 0.01$, and $p < 0.001$, respectively.

A detailed comparison shows that methods assuming an exponential distribution generally outperform those with Gamma or Poisson distributions. For instance, BABOR I-2 outperforms BABOR I-1 and I-3, BABOR II-2 surpasses BABOR II-1 and II-3, and BABOR III-2 exceeds BABOR III-1 and III-3. Among all methods, BABOR I-2 achieves the highest ASP (0.801) and ART (0.818), indicating superior performance. The advantage of the exponential distribution lies in two factors: (i) its effective modeling of continuous time data, compared to the discrete Poisson distribution, and (ii) its simplicity, requiring fewer parameters than the Gamma distribution, leading to better learning performance.

To assess the statistical significance of performance differences, a Wilcoxon signed-rank test was conducted on the average ASP and ART values between pairs of methods in Table 6. The p-values for BABOR I-1, I-2, and I-3 were all below 0.01, confirming that these models significantly outperform the baseline BOR model and the variants that only consider response time or attention on the criterion.

Table 7: Performance evaluation of methods across pairwise comparison tasks.

| Response Time Category | Index | Metric | BOR | BABOR I-1 | BABOR I-2 | BABOR I-3 | BABOR II-1 | BABOR II-2 | BABOR II-3 | BABOR III-1 | BABOR III-2 | BABOR III-3 |
|---|---|---|---|---|---|---|---|---|---|---|---|---|
| Long-Tail Distributions | $I_3$ | ASP | 0.465 | 0.658 | 0.656 | **0.664** | 0.561 | 0.629 | 0.517 | 0.499 | 0.648 | 0.483 |
|  |  | ART | 0.417 | 0.656 | 0.631 | **0.662** | 0.554 | 0.611 | 0.490 | 0.470 | 0.642 | 0.457 |
|  | $I_8$ | ASP | 0.413 | **0.756** | 0.749 | 0.753 | 0.666 | 0.725 | 0.535 | 0.554 | 0.738 | 0.454 |
|  |  | ART | 0.377 | **0.769** | 0.769 | 0.752 | 0.667 | 0.718 | 0.487 | 0.500 | 0.737 | 0.412 |
|  | $I_{28}$ | ASP | 0.665 | 0.823 | 0.819 | 0.813 | 0.735 | 0.834 | 0.692 | 0.752 | **0.845** | 0.686 |
|  |  | ART | 0.653 | 0.826 | 0.818 | 0.826 | 0.752 | 0.851 | 0.686 | 0.746 | **0.873** | 0.686 |
| Exceeding 20 Seconds | $I_4$ | ASP | 0.492 | 0.642 | **0.643** | 0.638 | 0.573 | 0.622 | 0.507 | 0.509 | 0.610 | 0.524 |
|  |  | ART | 0.472 | 0.634 | **0.641** | 0.626 | 0.537 | 0.626 | 0.511 | 0.504 | 0.606 | 0.520 |
|  | $I_{12}$ | ASP | 0.459 | 0.800 | **0.803** | 0.795 | 0.689 | 0.791 | 0.578 | 0.554 | 0.769 | 0.488 |
|  |  | ART | 0.410 | **0.787** | 0.787 | 0.787 | 0.694 | 0.769 | 0.565 | 0.505 | 0.762 | 0.438 |
| Less Than 20 Seconds | $I_{13}$ | ASP | 0.460 | 0.637 | **0.644** | 0.635 | 0.512 | 0.624 | 0.402 | 0.517 | 0.573 | 0.459 |
|  |  | ART | 0.414 | 0.639 | **0.656** | 0.648 | 0.525 | 0.631 | 0.393 | 0.517 | 0.569 | 0.474 |
|  | $I_{14}$ | ASP | 0.661 | 0.881 | **0.893** | 0.883 | 0.757 | 0.868 | 0.683 | 0.729 | 0.800 | 0.680 |
|  |  | ART | 0.650 | 0.886 | **0.895** | 0.895 | 0.781 | 0.876 | 0.695 | 0.718 | 0.806 | 0.660 |
|  | $I_{20}$ | ASP | 0.790 | 0.883 | **0.956** | 0.883 | 0.797 | 0.861 | 0.895 | 0.893 | 0.796 | 0.749 |
|  |  | ART | 0.850 | 0.867 | **0.960** | 0.867 | 0.810 | 0.857 | 0.905 | 0.900 | 0.800 | 0.810 |

Table 7 presents a detailed evaluation of various methods across pairwise comparison tasks characterized by distinct response time patterns, categorized in Section 4.2.1 into three groups: response time exceeding 20 seconds, less than 20 seconds, and long-tail distributions. Consistent with prior analyses, models integrating both response time and attention outperform single-factor models and the BOR baseline. Across the response time intervals, several trends emerge. Tasks with shorter response time (e.g., $I_{13}$, $I_{14}$ and $I_{20}$) exhibit more pronounced variation, with some methods achieving notably high or low ASP and ART, potentially reflecting either clear preferences enabling rapid decisions or rushed, less deliberate choices. For longer response time (e.g., $I_4$ and $I_{12}$), greater variability among methods is evident, with BABOR I-1 through I-3 excelling due to their adaptability to complex decision-making or cognitive fatigue. In tasks with long-tail distributions (e.g., $I_3$, $I_8$, and $I_{28}$), BABOR I-2 and III-2 demonstrate consistent robustness, achieving superior ASP and ART values. Despite task-specific differences, all BABOR variants consistently outperform the BOR baseline, highlighting the advantage of incorporating both response time and attention in predictive frameworks.

*4.2.3. Exploring heterogeneity in individual decision-making preferences*

This section provides a comprehensive analysis of decision-making preferences, beginning with an examination of overarching trends across all participants. Building on this foundation, we delve into individual-level heterogeneity, demonstrating how our approach captures unique decision behaviors. By integrating behavioral data, these analyses reveal nuanced patterns that enhance the model's explanatory power and predictive accuracy.

Figure 5a shows the relationship between response time and the posterior mean of comprehensive value difference under three distribution models. The overall trend indicates that response time decreases as the comprehensive value difference increases. This reflects the decision-making trade-off mechanism employed by individuals when evaluating different alternatives. Specifically, when the differences between alternatives are small, DMs are more likely to allocate additional time to analysis in order to reduce the risk of making errors. In contrast, when the differences are significant, DMs may adopt faster decision strategies, which can be interpreted as an optimization of cognitive resource allocation [20, 43]. Additionally, Figure 5b illustrates the relationship between attention duration and the posterior mean of marginal value differences under the same three distribution models. The results consistently demonstrate that attention duration decreases as marginal value differences increase. This suggests that substantial differences between alternatives on specific criteria reduce the cognitive effort required to process these differences, highlighting the role of value contrast in guiding attentional focus [33, 59].

We then analyze participants' decision-making patterns across tasks and criteria and employ the fuzzy C-means (FCM) clustering algorithm [10] to classify participants into subgroups. Unlike traditional hard

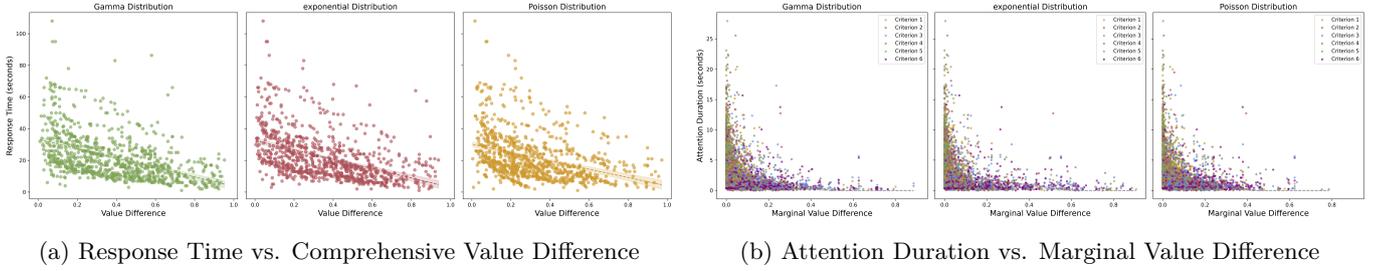

(a) Response Time vs. Comprehensive Value Difference     (b) Attention Duration vs. Marginal Value Difference

Figure 5: Relationships between value difference, response time, and attention duration across three distributions.

*Notes.* Figure 5a shows the relationship between response time and the posterior mean of comprehensive value difference across three distributions, based on 30 participants and 30 pairwise comparisons. Figure 5b illustrates the relationship between attention duration and the posterior mean of marginal value difference across the same three distributions, evaluated over 30 participants, 30 pairwise comparisons, and 6 criteria. Each point corresponds to a specific criterion.

clustering methods, such as K-means, which assign each data point to a single cluster, FCM offers a more nuanced perspective by allowing data points to belong to multiple clusters with varying degrees of membership.

*Analysis of decision-making efficiency across tasks.* We analyzed the decision-making efficiency of individuals across various tasks by examining their response times and group membership matrices (see Figure 6a). The results reveal significant heterogeneity in decision-making behaviors among individuals. For instance, individuals such as $d_1$ and $d_2$ exhibited stable behavior patterns throughout the experiment, which may indicate fixed preferences. Conversely, individuals such as $d_{16}$ and $d_{23}$ consistently spent more time across all tasks and criteria, suggesting a propensity for detailed deliberation and carefully weighing multiple criteria. In contrast, individuals such as $d_9$ and $d_{10}$ displayed notable variability in their response time across tasks, implying task-dependent shifts in preferences. To further validate these results, statistical analyses were conducted to compare the average response time between the identified groups (see Table 8). The results revealed significant differences in overall response time among the three groups, supporting the conclusion that their decision-making strategies differ systematically.

*Analysis of individual preferences across criteria.* Figure 6b illustrates the average attention duration allocated to different criteria, revealing a grouping pattern that closely aligns with the classifications derived from Figure 6a. Table 10 further evaluates the statistical significance of criterion importance across groups. The results indicate substantial inter-group differences in criterion preferences. Monthly fee ($g_5$) and domestic data ($g_2$) consistently emerge as the most influential criteria, exhibiting notable variation across groups, while initial deposit ($g_6$) is uniformly perceived as less important. These findings suggest that $g_5$ and $g_2$ are likely the primary drivers of decision-making. Figure 6c presents individual preference models, highlighting distinct group-level variations. Group 3 (indicated in purple) exhibits pronounced price sensitivity, prioritizing $g_5$ almost exclusively. In contrast, Group 4 (in pink) places greater emphasis on criteria related to excess consumption ($g_3$ and $g_4$), reflecting a more nuanced evaluation of cost structures. Group 1 (in green) demonstrates a more evenly distributed preference across $g_1$ through $g_6$. These patterns align with the findings in Figure 6b, where $g_5$ and $g_2$ are associated with higher attention durations and greater perceived importance, reinforcing the notion that individuals allocate more attention to criteria they consider critical.

*Analysis of perceived differences among individuals on specific criteria.* Figure 6d visualizes perceived differences across specific criteria in pairwise comparisons. When considered alongside Figure 6b, the findings support the hypothesis that criteria receiving greater attention (e.g., $g_2$ and $g_5$) tend to exhibit smaller perceived value differences across individuals, whereas lower-attention criteria display larger perceived differences. This pattern may be explained through the role of reference points [7]. For highly salient criteria, DMs may have established reference points or expectations, leading to minimal perceived differences even

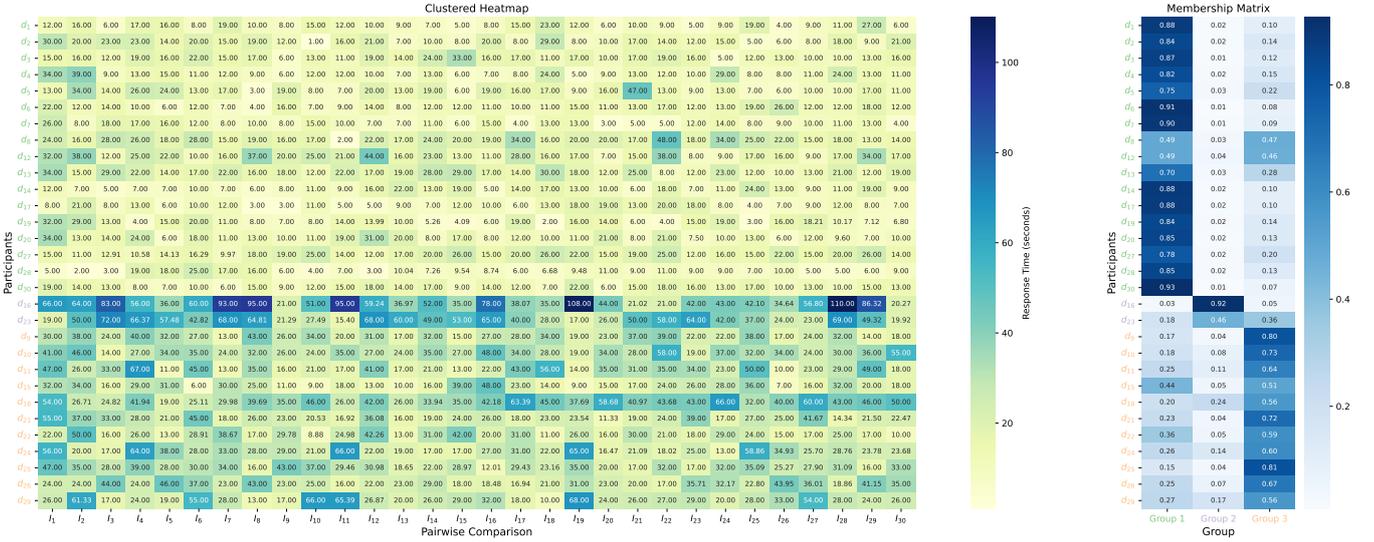

(a) Response time and group membership matrix for pairwise comparisons across participants.

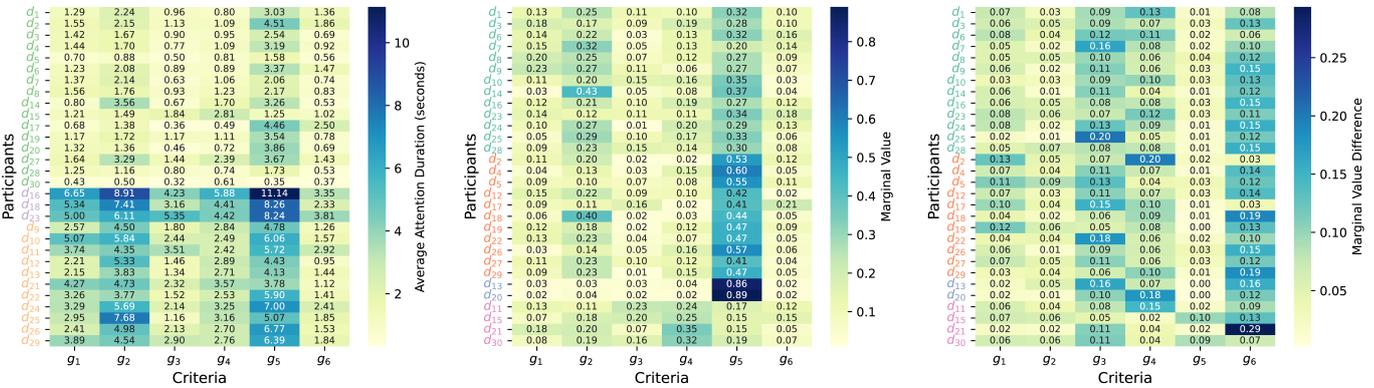

(b) Average attention duration     (c) Preference models     (d) Perceived criterion differences

Figure 6: Heterogeneity analysis of response time, attention duration, and preferences in decision-making process.

*Notes.* Participants were classified using the FCM algorithm, which assigns membership degrees indicating the extent of association with each cluster. Figure 6a shows response time and group membership matrices for pairwise comparisons, classifying 30 participants into three groups. Figure 6b illustrates average attention duration across criteria, with classifications consistent with Figure 6a. Figure 6c depicts preference models, dividing participants into four groups. Figure 6d highlights perceived differences across specific criteria for pairwise comparisons, aligned with the classifications in Figure 6c. The vertical axis labels in different colors represent participant groups, with the same color indicating members of the same group.

when changes are substantial but fall within the expected range. In contrast, for less salient criteria, the absence of clear reference points may heighten perceived differences, even when changes are relatively minor.

*Decision patterns analysis of participants.* By analyzing differences in response time patterns and criterion preferences, we further categorized participants into distinct groups and provided detailed characterizations of their features (see Table 11). Participants in Group 1 exhibited minimal variability in response time across tasks, with stable preferences and careful consideration of all criteria, suggesting a balanced and cautious decision-making pattern. Similarly, Group 2 participants also displayed stable preferences but prioritized price and functionality, reflecting a strong emphasis on cost-effectiveness and performance. Group 3 was characterized by extreme price sensitivity, suggesting that price was the sole dominant criterion in their decision-making process. Group 4 represented a distinct outlier group, with shorter response time but consistent and stable preferences, potentially indicating familiarity with specific task contexts or habitual responses. Group 5 had notably longer response time and comprehensively considered all criteria, reflecting high engagement and a meticulous decision-making approach. In contrast, Groups 6, 7, and 8 exhibited

Table 8: Statistical significance of average response time across various pairs (G1, G2) of participant groups.

| Pairwise comparison | (G1, G2) | Mean diff. | p-value | Pairwise comparison | (G1, G2) | Mean diff. | p-value | Pairwise comparison | (G1, G2) | Mean diff. | p-value |
|---|---|---|---|---|---|---|---|---|---|---|---|
| $I_1$ | (2,1) | 20.912 | 0.091 | $I_{11}$ | (2,1) | 43.140 | 0.004** | $I_{21}$ | (2,1) | 20.512 | 0.025* |
|  | (3,1) | 17.867 | 0.003** |  | (3,1) | 18.373 | 0.018* |  | (3,1) | 11.096 | 0.019* |
|  | (3,2) | -3.045 | 0.948 |  | (3,2) | -24.767 | 0.133 |  | (3,2) | -9.416 | 0.438 |
| $I_2$ | (2,1) | 38.706 | 0.000*** | $I_{12}$ | (2,1) | 46.974 | 0.000*** | $I_{22}$ | (2,1) | 21.853 | 0.069 |
|  | (3,1) | 17.892 | 0.001** |  | (3,1) | 13.735 | 0.003** |  | (3,1) | 13.144 | 0.031* |
|  | (3,2) | -20.814 | 0.061 |  | (3,2) | -33.240 | 0.000*** |  | (3,2) | -8.709 | 0.644 |
| $I_3$ | (2,1) | 63.741 | 0.000*** | $I_{13}$ | (2,1) | 35.601 | 0.000*** | $I_{23}$ | (2,1) | 41.323 | 0.000*** |
|  | (3,1) | 10.500 | 0.007** |  | (3,1) | 5.629 | 0.043* |  | (3,1) | 16.297 | 0.000*** |
|  | (3,2) | -53.244 | 0.000*** |  | (3,2) | -29.973 | 0.000*** |  | (3,2) | -25.026 | 0.000*** |
| $I_4$ | (2,1) | 44.505 | 0.000*** | $I_{14}$ | (2,1) | 35.882 | 0.000*** | $I_{24}$ | (2,1) | 27.794 | 0.004** |
|  | (3,1) | 20.586 | 0.000*** |  | (3,1) | 11.013 | 0.001** |  | (3,1) | 14.310 | 0.004** |
|  | (3,2) | -23.919 | 0.022* |  | (3,2) | -24.869 | 0.000*** |  | (3,2) | -13.484 | 0.234 |
| $I_5$ | (2,1) | 32.966 | 0.000*** | $I_{15}$ | (2,1) | 30.198 | 0.001** | $I_{25}$ | (2,1) | 26.906 | 0.001** |
|  | (3,1) | 12.773 | 0.002** |  | (3,1) | 12.377 | 0.006** |  | (3,1) | 22.239 | 0.000*** |
|  | (3,2) | -20.193 | 0.012* |  | (3,2) | -17.820 | 0.051 |  | (3,2) | -4.667 | 0.758 |
| $I_6$ | (2,1) | 35.980 | 0.000*** | $I_{16}$ | (2,1) | 59.162 | 0.000*** | $I_{26}$ | (2,1) | 17.321 | 0.044* |
|  | (3,1) | 17.482 | 0.000*** |  | (3,1) | 16.995 | 0.000*** |  | (3,1) | 13.922 | 0.002** |
|  | (3,2) | -18.497 | 0.042* |  | (3,2) | -42.167 | 0.000*** |  | (3,2) | -3.399 | 0.879 |
| $I_7$ | (2,1) | 67.560 | 0.000*** | $I_{17}$ | (2,1) | 24.741 | 0.009** | $I_{27}$ | (2,1) | 28.827 | 0.002** |
|  | (3,1) | 12.938 | 0.000*** |  | (3,1) | 16.230 | 0.001** |  | (3,1) | 20.682 | 0.000*** |
|  | (3,2) | -54.622 | 0.000*** |  | (3,2) | -8.511 | 0.531 |  | (3,2) | -8.145 | 0.564 |
| $I_8$ | (2,1) | 66.789 | 0.000*** | $I_{18}$ | (2,1) | 15.519 | 0.146 | $I_{28}$ | (2,1) | 75.984 | 0.000*** |
|  | (3,1) | 15.763 | 0.001** |  | (3,1) | 10.124 | 0.054 |  | (3,1) | 14.853 | 0.000*** |
|  | (3,2) | -51.026 | 0.000*** |  | (3,2) | -5.394 | 0.790 |  | (3,2) | -61.131 | 0.000*** |
| $I_9$ | (2,1) | 9.909 | 0.157 | $I_{19}$ | (2,1) | 48.060 | 0.003** | $I_{29}$ | (2,1) | 52.698 | 0.000*** |
|  | (3,1) | 14.108 | 0.000*** |  | (3,1) | 17.127 | 0.045* |  | (3,1) | 12.916 | 0.014* |
|  | (3,2) | 4.199 | 0.715 |  | (3,2) | -30.933 | 0.073 |  | (3,2) | -39.783 | 0.000*** |
| $I_{10}$ | (2,1) | 27.067 | 0.012* | $I_{20}$ | (2,1) | 23.236 | 0.007** | $I_{30}$ | (2,1) | 8.341 | 0.461 |
|  | (3,1) | 16.225 | 0.004** |  | (3,1) | 13.370 | 0.003** |  | (3,1) | 16.352 | 0.000*** |
|  | (3,2) | -10.843 | 0.460 |  | (3,2) | -9.866 | 0.371 |  | (3,2) | 8.010 | 0.508 |
| Overall comparison | (2,1) | - | 0.000*** | Overall comparison | (3,1) | - | 0.000*** | Overall comparison | (3,2) | - | 0.000*** |

*Notes.* *, ** and *** denote significance levels ANOVA and Tukey HSD post hoc tests, indicating significant differences between groups at $p < 0.05$, $p < 0.01$, and $p < 0.001$, respectively. *Mean diff.* represents the mean difference between group pairs, reflecting the magnitude of differences in the compared metrics.

Table 9: Statistical significance of average attention duration across criteria for various pairs (G1, G2) of participant groups.

| Criterion | (G1, G2) | Mean diff. | p-value | Criterion | (G1, G2) | Mean diff. | p-value |
|---|---|---|---|---|---|---|---|
| $g_1$ | (2,1) | 4.472 | 0.000*** | $g_4$ | (2,1) | 3.746 | 0.000*** |
|  | (3,1) | 2.065 | 0.000*** |  | (3,1) | 1.692 | 0.000*** |
|  | (3,2) | -2.407 | 0.000*** |  | (3,2) | -2.054 | 0.000*** |
| $g_2$ | (2,1) | 5.658 | 0.000*** | $g_5$ | (2,1) | 6.427 | 0.000*** |
|  | (3,1) | 3.205 | 0.000*** |  | (3,1) | 2.669 | 0.000*** |
|  | (3,2) | -2.453 | 0.002** |  | (3,2) | -3.757 | 0.000*** |
| $g_3$ | (2,1) | 3.387 | 0.000*** | $g_6$ | (2,1) | 2.145 | 0.000*** |
|  | (3,1) | 1.203 | 0.000*** |  | (3,1) | 0.645 | 0.025* |
|  | (3,2) | -2.184 | 0.000*** |  | (3,2) | -1.500 | 0.002** |

greater preference variability and significant differences in decision efficiency. Group 6 focused on external expenditures, Group 7 emphasized price and functionality, while Group 8, despite dynamic preference changes, demonstrated careful consideration of all criteria, indicative of integrative information processing in complex tasks. These findings highlight the heterogeneity in decision preferences and efficiency, providing empirical insights for personalized interventions and strategy optimization.

*Analysis of decision process for a single DM.* The results returned by BOR and BABOR variants consist of value-based preference models and the respective stochastic acceptabilities. In eAppendix C, we demonstrate the methods' application for a single DM ($d_{29}$). We discuss preference information, rank acceptability indices, distribution of feasible models, and the interrelations between behavioral cues, rankings, and criteria weights. This way, we exemplify the use of BOR and BABOR for real-world decision-aiding and show the impact of behavioral cues in the results.

Table 10: Statistical significance of criteria importance for various pairs (G1, G2) of participant groups.

| Criterion | (G1, G2) | Mean diff. | $p$-value | Criterion | (G1, G2) | Mean diff. | $p$-value |
|---|---|---|---|---|---|---|---|
| $g_1$ | (2,1) | -0.036 | 0.282 | $g_4$ | (2,1) | -0.044 | 0.135 |
| | (3,1) | -0.104 | 0.039* | | (3,1) | -0.107 | 0.032* |
| | (4,1) | -0.014 | 0.960 | | (4,1) | 0.153 | 0.000*** |
| | (3,2) | -0.068 | 0.274 | | (3,2) | -0.063 | 0.341 |
| | (4,2) | 0.022 | 0.853 | | (4,2) | 0.197 | 0.000*** |
| | (4,3) | 0.090 | 0.155 | | (4,3) | 0.260 | 0.000*** |
| $g_2$ | (2,1) | -0.051 | 0.353 | $g_5$ | (2,1) | 0.185 | 0.000*** |
| | (3,1) | -0.214 | 0.004** | | (3,1) | 0.574 | 0.000*** |
| | (4,1) | -0.079 | 0.275 | | (4,1) | -0.136 | 0.001** |
| | (3,2) | -0.162 | 0.040* | | (3,2) | 0.390 | 0.000*** |
| | (4,2) | -0.027 | 0.922 | | (4,2) | -0.321 | 0.000*** |
| | (4,3) | 0.135 | 0.180 | | (4,3) | -0.710 | 0.000*** |
| $g_3$ | (2,1) | -0.032 | 0.380 | $g_6$ | (2,1) | -0.023 | 0.656 |
| | (3,1) | -0.062 | 0.329 | | (3,1) | -0.080 | 0.148 |
| | (4,1) | 0.078 | 0.036* | | (4,1) | -0.003 | 0.999 |
| | (3,2) | -0.031 | 0.835 | | (3,2) | -0.057 | 0.418 |
| | (4,2) | 0.110 | 0.002** | | (4,2) | 0.020 | 0.886 |
| | (4,3) | 0.140 | 0.010* | | (4,3) | 0.078 | 0.264 |

Table 11: Decision-making profiles across participant groups.

| Group | Participants | Description |
|---|---|---|
| 1 | $d_1, d_6, d_7, d_8, d_{14}, d_{28}$ | Stable preferences, balanced consideration of all criteria, cautious decision-making. |
| 2 | $d_2, d_3, d_4, d_5, d_{12}, d_{17}, d_{19}, d_{27}$ | Stable preferences, focus on price and functionality, cost-effectiveness-driven. |
| 3 | $d_{13}, d_{20}$ | Extremely price-sensitive, price as the sole dominant criterion. |
| 4 | $d_{30}$ | Short response time, stable preferences, habitual or familiar task responses. |
| 5 | $d_{16}, d_{23}$ | Long response time, comprehensive consideration of all criteria, meticulous approach. |
| 6 | $d_{11}, d_{15}, d_{21}$ | Significant preference variability, focus on external expenditures. |
| 7 | $d_{18}, d_{29}, d_{22}, d_{26}$ | Significant preference variability, emphasis on price and functionality. |
| 8 | $d_9, d_{10}, d_{24}, d_{25}$ | Significant preference variability, broad consideration of all criteria, integrative decision-making. |

## 5. Conclusions

In this study, we proposed a novel preference learning approach that jointly integrates pairwise comparisons, response times, and attention durations within a Bayesian framework. Unlike traditional MCDA methods, which primarily rely on holistic judgments of alternatives and preference disaggregation for model calibration, our approach incorporates behavioral cues—specifically, response time and attention duration—directly into the preference elicitation process. By leveraging this auxiliary information, we extract additional insights into decision-making patterns, enhancing the construction of the preference model. To validate the effectiveness and robustness of our approach, we conducted a laboratory experiment on contract selection, demonstrating its applicability and shedding light on the influence of behavioral cues in preference formation.

We recruited 30 Chinese students from Xi'an Jiaotong University to participate in a laboratory experiment. Participants performed pairwise comparisons of 10 real-world mobile phone contracts offered by three major Chinese network operators. Throughout the experiment, we captured multiple behavioral data, including gaze duration for each alternative, attention duration for each criterion, response time for each pairwise comparison, and the coordinates of mouse cursor movements. Through ablation studies and comparative analyses, we demonstrated that our proposed approach significantly outperforms alternative model variants—specifically, those that consider only criteria-level attention, those that incorporate only response time, and the baseline BOR method. These results underscore the importance of integrating multiple behavioral data to develop preference models that more accurately reflect DMs' true preferences. Furthermore, we examined the relationships between response time and value differences, as well as between attention duration and the posterior mean of marginal value differences. The results indicate that response time decreases as value differences increase, while attention duration also declines as marginal value differences become more pronounced. These findings suggest that DMs employ a systematic trade-off strategy when evaluating alternatives. Specifically, when the differences between alternatives are small, DMs tend to spend more time deliberating to minimize the risk of errors. Conversely, when the differences are substantial, faster decision-making strategies are adopted, reflecting an optimization of cognitive resources. This evidence supports the theoretical assumptions underlying our model and aligns with established decision-making frameworks [20, 43].

We conducted a detailed analysis of participants' decision-making patterns across tasks and criteria to explore further the role of behavioral cues in understanding individual decision processes. By assessing whether differences in response time across tasks were statistically significant, we identified distinct decision-making profiles, including deliberative DMs who allocate extended time to carefully weighing multiple criteria, consistent DMs with specific preferences, and task-dependent DMs whose response times fluctuate depending on the complexity of the task. Moreover, we observed that tasks with longer response times often involved smaller differences between alternatives, suggesting that participants engaged in deeper cognitive processing to make their choices. Additionally, attention duration on criteria provided insights into participants' preferences, as most participants allocated more attention to the criteria they deemed most important.

These findings highlight the critical role of behavioral cues in understanding individual preferences, enriching preference profiling, and offering actionable insights for personalized interventions and differentiated marketing strategies. For individuals with balanced preferences and cautious decision-making styles, emphasizing holistic product value and long-term incentives may enhance confidence in making stable choices. For those prioritizing cost-efficiency and functionality, recommending high-value products with tiered service options may address their dual concerns about price and functionality. Price-sensitive individuals may respond more favorably to low-cost strategies and time-limited promotions, particularly when coupled with clear and transparent pricing information to encourage swift decision-making. Finally, for context-dependent individuals, implementing dynamic pricing and tailored promotions can enhance engagement and satisfaction by aligning offerings with situational preferences.

We envisage the following directions for future research. First, the proposed method can be readily adapted to address multiple criteria sorting problems, multiple criteria hierarchy process [3], scenarios with interacting criteria [6], and contexts involving potentially non-monotonic marginal value functions [56]. Extensions of the method to other score- [3] or outranking-based [24] models, where Stochastic Ordinal Regression and Stochastic Multicriteria Acceptability Analysis have previously been applied [51], also represent natural avenues for exploration. Second, integrating additional process-tracing data, such as saccades and fixation counts [76] or heart rate and EEG signals, offers an intriguing opportunity to elucidate further the underlying cognitive and psychological mechanisms in the MCDA domain. Third, it would be possible to develop adaptive preference elicitation techniques that adjust the process in real time based on observed behavioral cues (e.g., increasing granularity when response time suggests uncertainty). Fourth, refining the experimental design by incorporating more rigorous participant recruitment strategies across diverse backgrounds and developing context-specific scenarios could improve the robustness and generalizability of findings. Fifth, developing explanatory and visualization techniques that help DMs understand how behavioral data influences the ranking of alternatives is worthwhile. Finally, evaluating the proposed method's performance and robustness in solving complex, real-world decision-making problems remains an essential direction. This particularly appeals to high-stakes decision problems, such as medical diagnosis, financial investment, or policy-making, where attention allocation and response time may vary substantially from one pair to another.

**Acknowledgment**

Jiapeng Liu acknowledges support from the National Natural Science Foundation of China (grant no. 72471184, 72071155). Miłosz Kadziński was supported by the Polish National Science Center under the SONATA BIS project (grant no. DEC-2019/34/E/HS4/00045).

## Appendix A. Approximating value function model using piecewise-linear functions

In this appendix, we present the details of approximating value function model $U(\cdot)$ using piecewise-linear functions. Given a sufficient number of subintervals, such functions can approximate any nonlinear shape [1]. This property enhances the flexibility and the expressiveness of an assumed model in representing the DM's sophisticated preferences [3]. Moreover, the shape of a piecewise-linear marginal value function reflects the DM's risk attitude [2]. Also, using such functions is computationally tractable in a probabilistic inference procedure [4].

Specifically, piece-wise linear functions with a limited number of characteristic points are particularly useful for practical decision making. The evaluation scale $X_j = [x_{j_*}, x_{j^*}]$ of criterion $g_m$ is divided into $\gamma_j \geqslant 1$ equal-length subintervals $[x_j^0, x_j^1], [x_j^1, x_j^2], \ldots, [x_j^{\gamma_j-1}, x_j^{\gamma_j}]$, where $x_{j_*}$ is the worst performance on $g_m$, $x_{j^*}$ is the best performance and $x_j^k = x_{j_*} + \frac{k}{\gamma_j}(x_{j^*} - x_{j_*}), k = 0, 1, \ldots, \gamma_j$. Then, the marginal value function of $a$ on $g_m$ can be calculated as follows:

$$U_m(g_m(a_n)) = U_m(x_j^{k_j}) + \frac{g_m(a_n) - x_j^{k_j}}{x_j^{k_j+1} - x_j^{k_j}} \left( U_m(x_j^{k_j+1}) - U_m(x_j^{k_j}) \right), \text{ for } g_m(a_n) \in \left[ x_j^{k_j}, x_j^{k_j+1} \right]. \quad (A.1)$$

According to the above formula, the marginal value $U_m(g_m(a_n))$ is estimated through linear interpolation. Once the marginal values at knots are determined, the marginal value function $U_m(\cdot)$ can be fully specified.

We now introduce a new variable $\Delta U_m^t$ to denote the difference between marginal values at two consecutive characteristic points $x_j^t$ and $x_j^{t-1}$, $\Delta U_m^t = U_m(x_j^t) - U_m(x_j^{t-1})$, $t = 1, \ldots, \gamma_j$. $\Delta U_m^t$ could be positive or negative, which implies the marginal value function of increasing or decreasing preference within the interval $\left[x_j^{t-1}, x_j^t\right]$. Then, we can rewrite the marginal value function $U_m(g_m(a_n))$ correspondingly:

$$U_m(g_m(a_n)) = \sum_{t=1}^{k_j} \Delta U_m^t + \frac{g_m(a_n) - x_j^{k_j}}{x_j^{k_j+1} - x_j^{k_j}} \Delta U_m^{k_j+1}, \text{ for } g_m(a_n) \in \left[ x_j^{k_j}, x_j^{k_j+1} \right]. \quad (A.2)$$

We can express the marginal value $U_m(g_m(a_n))$ as an inner product between two vectors $\boldsymbol{u}_j$ and $\boldsymbol{V}_m(a_n)$, that is $U_m(g_m(a_n)) = \boldsymbol{u}_j^\top \boldsymbol{V}_m(a_n)$, where $\boldsymbol{u}_j = (\Delta U_m^1, \ldots, \Delta U_m^{\gamma_j})^\top$, and $\boldsymbol{V}_m(a_n) = (v_j^1(a_n), \ldots, v_j^{\gamma_j}(a_n))^\top$. When $U_m(g_m(a_n))$ is monotone increasing, for each $t = 1, 2, \ldots, \gamma_j$:

$$v_j^t(a_n) = \begin{cases} 1, & \text{if } g_m(a_n) > x_j^t, \\ \frac{g_m(a_n) - x_j^{t-1}}{x_j^t - x_j^{t-1}}, & \text{if } x_j^{t-1} \leqslant g_m(a_n) \leqslant x_j^t, \\ 0, & \text{otherwise.} \end{cases} \quad (A.3)$$

otherwise, $v_j^t(a_n)$ can be shown below when $U_m(g_m(a_n))$ is monotone decreasing respectively:

$$v_j^{\gamma_j - t}(a_n) = \begin{cases} 0, & \text{if } g_m(a_n) > x_j^t, \\ \frac{x_j^t - g_m(a_n)}{x_j^t - x_j^{t-1}}, & \text{if } x_j^{t-1} \leqslant g_m(a_n) \leqslant x_j^t. \\ 1, & \text{otherwise.} \end{cases} \quad (A.4)$$

Let us gather all $\boldsymbol{u}_j$ and $\boldsymbol{V}_m(a_n)$ to denote $\boldsymbol{u} = (\boldsymbol{u}_1^\top, \ldots, \boldsymbol{u}_m^\top)^\top$, and $\boldsymbol{V}(a_n) = (\boldsymbol{V}_1(a_n)^\top, \ldots, \boldsymbol{V}_m(a_n)^\top)^\top$. Particularly, we call $\boldsymbol{V}(a_n)$ the characteristic vector of alternative $a_n$, since it solely depends on the performances of $a_n$ on multiple criteria, whereas $\boldsymbol{u}$ decides upon the intrinsic character of the additive value model $U(a_n)$. Then the comprehensive value $U(a_n)$ of $a_n$ can be expressed as follows:

$$U(a_n) = \boldsymbol{u}^\top \boldsymbol{V}(a_n). \tag{A.5}$$

To ensure the monotonicity (non-decreasing / non-increasing) and normalization properties of the value function model $U$, it's essential to account for the following linera constraints:

$$\begin{cases} \boldsymbol{u}^\top \boldsymbol{1} = 1, \\ \boldsymbol{u} \geqslant \boldsymbol{0}, \end{cases} \tag{A.6}$$

where $\boldsymbol{1}$ and $\boldsymbol{0}$ are the vectors with all entries being equal to 1 and 0, respectively.

**Appendix B. Performance evaluation metrics**

To compare the performance of BOR and our nine model variants that consider different forms of response time and attention duration, we use the following two measures. The first is called *Average Support of the inferred Pairwise Winning Indices on the true pairwise comparisons between alternatives* (ASP). It quantifies the proportion of pairwise comparisons for which the inferred pairwise winning indices (PWI) are consistent with the true preference model. Specifically, it computes the average level of agreement between the inferred and true pairwise preferences across all alternatives. ASP is computed as:

$$\text{ASP} = \frac{1}{|I^T|} \sum_{(a_{s_1^l}, a_{s_2^l}) \in I^T} \text{PWI}(a_{s_1^l}, a_{s_2^l}),$$

where $I^T = \{I_1, \ldots, I_l, \ldots, I_S\}$ denotes the set of pairwise comparisons in the test set, and $|I^T| = S$ denotes the total number of pairwise comparisons in $I^T$. Each $I_l$ specifies a preference relation between alternatives $(a_{s_1^l}, a_{s_2^l})$, denoted as $a_{s_1^l} \succ a_{s_2^l}$. The term $\text{PWI}(I_l) \approx \frac{\left|S_{U(a_{s_1^l}) > U(a_{s_2^l})}\right|}{K}$, where $\left|S_{U(a_{s_1^l}) > U(a_{s_2^l})}\right|$ represents the number of sampled value functions for which $U(a_{s_1^l}) > U(a_{s_2^l})$. A higher value of ASP indicates a better alignment between the model's inferred preference and the true preference.

The second measure is named *Accuracy Rate on Test set* (ART). It evaluates the accuracy of the inferred pairwise comparisons against the true preference. It calculates the proportion of pairwise comparisons where the inferred preference matches the true preference. ART is calculated as:

$$\text{ART} = \frac{1}{|I^T|} \sum_{(a_{s_1^l}, a_{s_2^l}) \in I^T} \mathbb{I}(\text{PWI}(a_{s_1^l}, a_{s_2^l}) > 0.5),$$

where the indicator function $\mathbb{I}(\text{PWI}(a_{s_1^l}, a_{s_2^l}) > 0.5)$ takes the value 1 if $PWI(a_{s_1^l}, a_{s_2^l}) > 0.5$, and 0 otherwise. A higher value of ART indicates better performance in accurately predicting the true pairwise preferences.

**Appendix C. An illustrative example of BOR and BABOR for real-world decision aiding**

In this appendix, we demonstrate the application of the proposed approach by inferring a DM's preference model based on a real experimental decision-making scenario outlined in Section 4. We focus on the individual DM rather than all considered subjects, the respective preference information and model, and the relations between behavioral cues and the obtained results.

*Appendix C.1. Implementation of the proposed approach*

We apply BABOR to this case study and compare the results with those derived from BOR. The proposed method relies on preference information regarding reference alternatives to derive the posterior distribution of the value function. Preference data is collected using the experimental setup described in Section 4. Table C.1 shows the preference information used by the model, with data from object $d_{29}$ as an example. The shortest response time among the 30 pairwise comparisons is observed for $I_{18}$ ($a_2 \succ a_{10}$) at 10 seconds, while the longest response time occurs for $I_{19}$ ($a_1 \succ a_9$) at 68 seconds. The comparison with the greatest differentiation in attention duration across criteria is $I_{11}$ ($a_6 \succ a_8$), where attention durations range significantly from 2.398 seconds for $g_6$ to 22.397 seconds for $g_2$. On the other hand, the most balanced attention durations appear in $I_1$ ($a_8 \succ a_3$), where the values vary only slightly, from 2.332 seconds for $g_4$ to 3.980 seconds for $g_2$, showing the smallest disparity among all comparisons. The criterion with the least average attention duration is $g_6$, while the criterion with the most significant average attention duration is $g_5$. Note that, due to distinct model assumptions, only pairwise comparison information is utilized for BOR. On the contrary, the observed differences in response times and attention durations give the variants of BABOR means for inferring models that are better adjusted to the DM's indirect preferences and behavioral cues.

Table C.1: Preference information provided by Decision Maker $d_{29}$.

| Index | Pairwise comparison | Response time | Attention duration | Index | Pairwise comparison | Response time | Attention duration on each criterion |
|---|---|---|---|---|---|---|---|
| $I_1$ | $a_8 \succ a_3$ | 26.001 | (3.511, 3.980, 3.364, 2.332, 3.431, 3.231) | $I_{16}$ | $a_8 \succ a_4$ | 32.000 | (3.131, 9.260, 2.215, 3.914, 7.698, 1.266) |
| $I_2$ | $a_2 \succ a_9$ | 61.333 | (13.107, 7.195, 7.911, 13.074, 16.615, 3.431) | $I_{17}$ | $a_1 \succ a_4$ | 18.000 | (1.782, 5.829, 1.182, 1.133, 5.517, 2.515) |
| $I_3$ | $a_7 \succ a_3$ | 17.000 | (2.382, 1.649, 1.566, 1.832, 4.164, 0.566) | $I_{18}$ | $a_2 \succ a_{10}$ | 10.000 | (0.799, 1.382, 0.716, 0.766, 2.681, 0.799) |
| $I_4$ | $a_8 \succ a_5$ | 24.000 | (2.881, 1.066, 2.298, 0.533, 4.170, 1.416) | $I_{19}$ | $a_1 \succ a_9$ | 68.000 | (11.858, 6.111, 2.598, 3.981, 7.458, 1.836) |
| $I_5$ | $a_6 \succ a_4$ | 19.000 | (2.448, 3.581, 3.497, 0.933, 6.668, 0.300) | $I_{20}$ | $a_2 \succ a_8$ | 24.001 | (2.482, 3.548, 2.532, 3.864, 7.709, 2.132) |
| $I_6$ | $a_7 \succ a_9$ | 54.999 | (3.781, 3.031, 2.365, 4.280, 6.246, 1.316) | $I_{21}$ | $a_6 \succ a_9$ | 26.000 | (3.664, 5.047, 1.982, 2.548, 11.330, 1.132) |
| $I_7$ | $a_8 \succ a_1$ | 28.000 | (6.695, 3.364, 4.996, 4.613, 6.229, 1.616) | $I_{22}$ | $a_7 \succ a_4$ | 27.001 | (2.149, 6.423, 2.298, 1.466, 8.659, 0.966) |
| $I_8$ | $a_8 \succ a_3$ | 12.999 | (1.732, 2.232, 1.266, 2.165, 1.399, 0.933) | $I_{23}$ | $a_8 \succ a_9$ | 29.001 | (2.265, 2.762, 2.265, 0.616, 4.197, 2.048) |
| $I_9$ | $a_5 \succ a_{10}$ | 17.001 | (0.333, 1.099, 1.166, 0.666, 3.531, 0.450) | $I_{24}$ | $a_6 \succ a_2$ | 19.999 | (1.732, 4.336, 1.166, 1.182, 7.065, 1.282) |
| $I_{10}$ | $a_1 \succ a_{10}$ | 66.000 | (1.095, 2.598, 0.783, 0.316, 3.581, 0.683) | $I_{25}$ | $a_5 \succ a_7$ | 28.000 | (4.680, 3.048, 3.281, 2.315, 6.262, 2.082) |
| $I_{11}$ | $a_6 \succ a_8$ | 65.387 | (10.776, 22.397, 3.481, 4.164, 22.171, 2.398) | $I_{26}$ | $a_3 \succ a_{10}$ | 33.000 | (2.071, 3.248, 1.665, 0.283, 3.681, 1.616) |
| $I_{12}$ | $a_1 \succ a_3$ | 26.871 | (3.847, 3.691, 7.462, 6.046, 3.677, 2.148) | $I_{27}$ | $a_2 \succ a_7$ | 54.000 | (12.613, 8.244, 7.361, 5.546, 9.876, 8.977) |
| $I_{13}$ | $a_5 \succ a_4$ | 20.000 | (1.641, 1.699, 1.066, 1.466, 2.406, 0.833) | $I_{28}$ | $a_6 \succ a_{10}$ | 28.000 | (4.847, 2.831, 4.847, 4.780, 4.230, 4.930) |
| $I_{14}$ | $a_2 \succ a_3$ | 26.000 | (3.563, 5.063, 2.065, 1.999, 9.777, 2.365) | $I_{29}$ | $a_9 \succ a_4$ | 24.000 | (1.216, 2.015, 2.598, 1.782, 2.898, 0.716) |
| $I_{15}$ | $a_7 \succ a_1$ | 29.000 | (1.982, 5.180, 1.949, 2.715, 4.766, 0.550) | $I_{30}$ | $a_2 \succ a_1$ | 26.000 | (1.532, 4.166, 4.963, 1.416, 3.631, 0.550) |

From the Bayesian inference perspective, these three types of preference information are used to construct the likelihood of the value function. To implement our method, the following parameters must be established: (a) the numbers of sub-intervals $\gamma_m$ for each criterion $g_m$ ($m = 1, \ldots, M$), (b) the number of samples $K$ for approximating the posterior distribution, and (c) the number of random walks $W$ for the Markov chain to reach its stationary distribution. For simplicity, we assume all criteria share the same number of sub-intervals. The optimal value of $\gamma_m = 2$ was determined using cross-validation on the validation set. Then, we set $K = 10,000$, which is sufficient for accurate approximation [7], and $W = 1000$, based on convergence analysis.

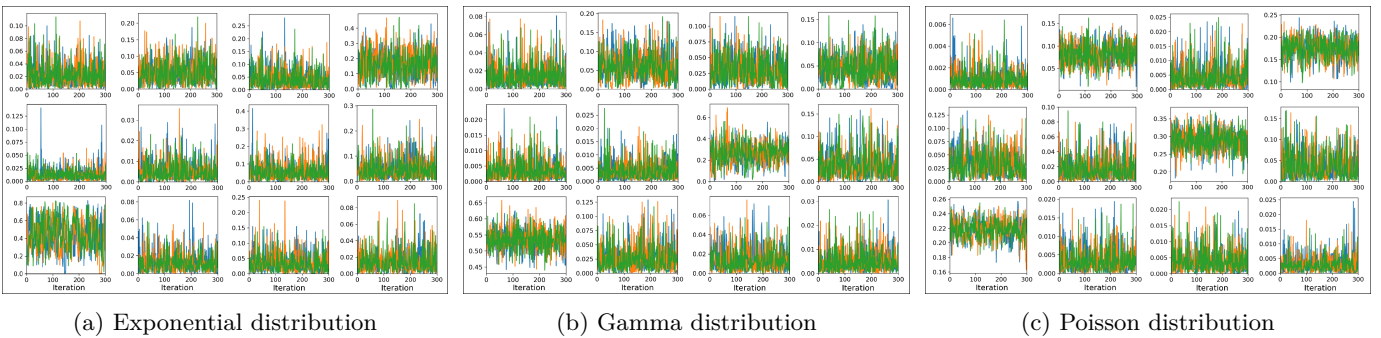

(a) Exponential distribution     (b) Gamma distribution     (c) Poisson distribution

Figure C.1: The trace plots for each entry of parameter $\boldsymbol{u}$ in the considered model assuming three different distributions for response times and attention durations.

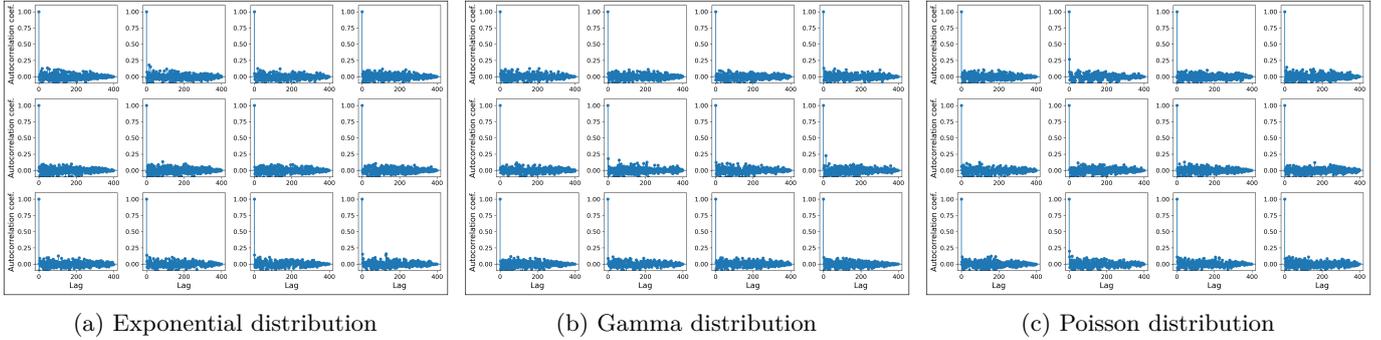

(a) Exponential distribution  (b) Gamma distribution  (c) Poisson distribution

Figure C.2: The autocorrelation coefficients for each entry of parameter $\boldsymbol{u}$ in the considered model for different distributions for response times and attention durations.

Using these parameter settings, we applied the proposed sampling procedure to obtain samples from the posterior distribution. After 11,000 iterations, we discarded the first 1,000 samples as burn-in, leaving 10,000 samples for posterior approximation. Figures C.1a–C.1c present trace plots for each entry of **u** under three distinct models that account for different forms of response time and attention duration. To evaluate the convergence of the Markov chains, we constructed three separate chains with distinct starting points, each visualized in a different color. The absence of any discernible trend in the trace plots indicates that the chains have effectively "forgotten" their initial states, confirming convergence [5]. Figures C.2a–C.2c illustrate the autocorrelation coefficients for each entry of **u** across the same model variants. For all entries, the autocorrelation coefficients rapidly decay to near zero, further confirming convergence to the stationary distribution. The rapid reduction in autocorrelation is a key indicator of convergence. By examining both the trace plots and autocorrelation coefficients for each model variant, we gain insights into the robustness and reliability of the sampling procedure across varying modeling assumptions.

*Appendix C.2. Result analysis*

We compare the results obtained by three variants of BABOR and BOR. Table C.2 presents the Rank Acceptability Indices (RAIs) for the ten alternatives. In all cases, alternative $a_2$ overwhelmingly attains the first rank, with indices of 97.77% in BOR, 100% in BABOR I-1 and I-2, and 99.94% in BABOR I-3, which clearly confirms its top position. The BABOR approaches yield more concentrated rank distributions compared to the more dispersed results observed with BOR. In terms of the most frequently attained ranks, a consistent pattern emerges: $a_6$ is predominantly ranked second (e.g., 85.42% in BOR and 100% in BABOR variants), while $a_3$ and $a_4$ are generally relegated to the lower ranks (mostly ninth and tenth, respectively, though with slightly lower RAIs for BOR than for BABOR variants). For alternatives $a_1$, $a_7$, and $a_9$, BOR produces rankings that differ from those of BABOR variants. Specifically, the latter ones rank $a_9$ in the sixth position with probabilities of 86.17%, 87.56%, and 64.99%, respectively, whereas BOR ranks $a_9$ in the fourth position with 44.35% probability. These differences arise because BABOR variants incorporate more types of preference information than BOR, allowing for more precise likelihood estimation, reducing uncertainty, and resulting in a more concentrated posterior distribution, yielding more concentrated RAIs.

We utilized the collected instances of the preference model $\{\mathbf{u}(t)\}_{t=W+1}^{K}$ to estimate the posterior means and 95% Highest Posterior Density (HPD) intervals for the marginal values at characteristic points (see Table C.3). The definition of HPD follows [6]. Given that $\gamma_j = 2$ is the optimal setting for the number of sub-intervals per criterion, each marginal value function consists of two pieces. Our analysis focuses on the results of BABOR I-2, which demonstrated the best performance, while the other BABOR variants yield similar findings and are omitted for brevity. The results indicate that the average trade-off weights across all criteria are 0.087, 0.234, 0.012, 0.151, 0.472, and 0.048. Additionally, the estimation uncertainty for marginal values at characteristic points is lowest for $g_1$ and $g_3$, as evidenced by the shorter average lengths of the HPD intervals.

Table C.2: The Rank Acceptability Indices (in %) derived from BOR and three variants of BABOR.

| Method | Rank | 1 | 2 | 3 | 4 | 5 | 6 | 7 | 8 | 9 | 10 |
|---|---|---|---|---|---|---|---|---|---|---|---|
| | $a_1$ | | 0.06 | | 0.34 | 9.96 | **88.55** | 0.56 | 0.40 | 0.08 | |
| | $a_2$ | **97.77** | 1.90 | 0.21 | | | | | | | |
| | $a_3$ | | | | | 0.05 | 0.15 | 1.57 | 4.54 | **93.62** | |
| | $a_4$ | | | | | 0.05 | 0.18 | 0.99 | 1.48 | 1.33 | **95.91** |
| BOR | $a_5$ | | | | | 0.07 | 0.19 | 9.31 | **83.77** | 3.16 | 3.42 |
| | $a_6$ | 0.76 | **85.42** | 9.82 | 3.93 | | | | | | |
| | $a_7$ | | | 5.04 | 38.96 | **49.35** | 0.95 | 0.19 | 0.10 | | |
| | $a_8$ | 0.19 | 10.65 | **74.26** | 12.17 | 1.90 | 0.57 | 0.11 | 0.10 | | |
| | $a_9$ | 1.28 | 1.91 | 10.58 | **44.35** | 38.36 | 8.89 | | | | |
| | $a_{10}$ | | | | 0.10 | 0.19 | 0.50 | **87.23** | 9.58 | 1.74 | 0.59 |
| Method | Rank | 1 | 2 | 3 | 4 | 5 | 6 | 7 | 8 | 9 | 10 |
| | $a_1$ | | | | | **86.17** | 13.83 | | | | |
| | $a_2$ | **100.00** | | | | | | | | | |
| | $a_3$ | | | | | | | 0.25 | 4.11 | **95.22** | 0.42 |
| | $a_4$ | | | | | | | | 0.37 | 0.49 | **99.10** |
| BABOR I-1 | $a_5$ | | | | | | | 0.88 | **94.51** | 4.13 | 0.48 |
| | $a_6$ | | **100.00** | | | | | | | | |
| | $a_7$ | | | 13.31 | **86.69** | | | | | | |
| | $a_8$ | | | **86.69** | 13.31 | | | | | | |
| | $a_9$ | | | | | 13.83 | **86.17** | | | | |
| | $a_{10}$ | | | | | | | **98.83** | 1.01 | 0.16 | |
| Method | Rank | 1 | 2 | 3 | 4 | 5 | 6 | 7 | 8 | 9 | 10 |
| | $a_1$ | | | | | **86.46** | 12.20 | 1.00 | | | |
| | $a_2$ | **100.00** | | | | | | | | | |
| | $a_3$ | | | | | | | | 0.86 | **99.06** | 0.08 |
| | $a_4$ | | | | | | | | 0.38 | 0.33 | **99.29** |
| BABOR I-2 | $a_5$ | | | | | | | 0.56 | **98.25** | 0.61 | 0.54 |
| | $a_6$ | | **100.00** | | | | | | | | |
| | $a_7$ | | | 6.21 | **93.79** | | | | | | |
| | $a_8$ | | | **93.79** | 6.21 | | 0.22 | | | | |
| | $a_9$ | | | | | 9.59 | **87.56** | 2.68 | 0.19 | | |
| | $a_{10}$ | | | | | 3.94 | | **95.80** | 0.30 | | |
| Method | Rank | 1 | 2 | 3 | 4 | 5 | 6 | 7 | 8 | 9 | 10 |
| | $a_1$ | | | | | **86.20** | 13.71 | 0.05 | | | |
| | $a_2$ | **99.94** | 0.06 | | | | | | | | |
| | $a_3$ | | | | | | | 0.87 | 2.20 | **95.17** | 1.71 |
| | $a_4$ | | | | | | | 0.37 | 0.73 | 1.52 | **97.34** |
| BABOR I-3 | $a_5$ | | | | | | | 8.36 | **88.42** | 2.58 | 0.63 |
| | $a_6$ | 0.06 | **99.94** | | | | | | | | |
| | $a_7$ | | | 22.42 | **56.32** | 0.10 | 21.24 | | | | |
| | $a_8$ | | | **77.20** | 22.76 | | | | | | |
| | $a_9$ | | | 0.37 | 20.92 | 13.70 | **64.99** | | | | |
| | $a_{10}$ | | | | | | | **90.33** | 8.60 | 0.73 | 0.32 |

Table C.3: Posterior means and 95% HPD Intervals of marginal values at characteristic points in BABORs methods.

| Criterion | Model | Marginal value | | | | Criterion | Model | Marginal value | | | |
|---|---|---|---|---|---|---|---|---|---|---|---|
| | | Posterior mean | | HPD interval | | | | Posterior mean | | HPD interval | |
| | | $\Delta U_m^1$ | $\Delta U_m^2$ | $[\Delta U_{mL}^1, \Delta U_{mR}^1]$ | $[\Delta U_{mL}^2, \Delta U_{mR}^2]$ | | | $\Delta U_m^1$ | $\Delta U_m^2$ | $[\Delta U_{mL}^1, \Delta U_{mR}^1]$ | $[\Delta U_{mL}^2, \Delta U_{mR}^2]$ |
| | BABOR I-1 | 0.009 | 0.069 | [0,0.028] | [0, 0.180] | | BABOR I-1 | 0.216 | 0.028 | [0.024, 0.391] | [0, 0.081] |
| $g_1$ | BABOR I-2 | 0.019 | 0.068 | [0, 0.020] | [0, 0.074] | $g_4$ | BABOR I-2 | 0.082 | 0.068 | [0, 0.242] | [0, 0.201] |
| | BABOR I-3 | 0.001 | 0.080 | [0, 0.004] | [0, 0.219] | | BABOR I-3 | 0.279 | 0.049 | [0, 0.518] | [0, 0.115] |
| | BABOR I-1 | 0.039 | 0.048 | [0, 0.110] | [0, 0.136] | | BABOR I-1 | 0.520 | 0.026 | [0.293, 0.748] | [0, 0.075] |
| $g_2$ | BABOR I-2 | 0.036 | 0.198 | [0, 0.057] | [0, 0.396] | $g_5$ | BABOR I-2 | 0.456 | 0.017 | [0.384, 0.875] | [0, 0.050] |
| | BABOR I-3 | 0.002 | 0.183 | [0, 0.006] | [0,0.408] | | BABOR I-3 | 0.228 | 0.003 | [0, 0.560] | [0, 0.008] |
| | BABOR I-1 | 0.004 | 0.004 | [0, 0.012] | [0, 0.012] | | BABOR I-1 | 0.022 | 0.014 | [0,0.066] | [0, 0.043] |
| $g_3$ | BABOR I-2 | 0.007 | 0.005 | [0, 0.021] | [0,0.015] | $g_6$ | BABOR I-2 | 0.033 | 0.015 | [0, 0.094] | [0, 0.045] |
| | BABOR I-3 | 0.041 | 0.018 | [0,0.123] | [0, 0.053] | | BABOR I-3 | 0.003 | 0.002 | [0, 0.009] | [0, 0.007] |

*Notes.* $[\Delta U_{mL}^{\gamma_j}, \Delta U_{mR}^{\gamma_j}]$ represents the 95% Highest Posterior Density (HPD) interval of the marginal value on criterion $g_m$ at characteristic point $\gamma_j$, as defined in [6], quantifying the uncertainty of parameter $\Delta U_m^{\gamma_j}$.

We visualized the decision process of $d_{29}$ for all pairwise comparisons in Figure C.3. The DM exhibited notably longer response time for certain tasks, including $I_2$, $I_6$, $I_{10}$, $I_{11}$, $I_{19}$, and $I_{27}$. Cross-referencing these findings with RAIs in Table C.2, we observe that longer response times are associated with comparisons involving adjacent rankings, such as $(a_1, a_9)$, $(a_6, a_8)$, $(a_1, a_{10})$, and $(a_7, a_9)$. Additionally, response times are also prolonged for comparisons involving the preferred alternative $a_2$. These patterns suggest that response time can serve as an informative cue for assessing the distinctiveness between alternatives and inferring a DM's preferences.

Furthermore, we estimated the relative importance of criteria based on attention duration. The observed ranking of importance, derived from the average attention durations for each criterion, is $g_5$, $g_2$, $g_1$, $g_3$, $g_4$, and $g_6$. Meanwhile, the ranking based on the maximal shares in the comprehensive values for the preference model is $g_5$, $g_2$, $g_4$, $g_1$, $g_6$, and $g_3$. The consistency between these two rankings suggests that attention allocation reasonably approximates the impacts of various criteria on the suggested recommendation. Overall, our findings indicate that $d_{29}$ places particular emphasis on price and functionality, suggesting that these criteria play a dominant role in her decision-making process.

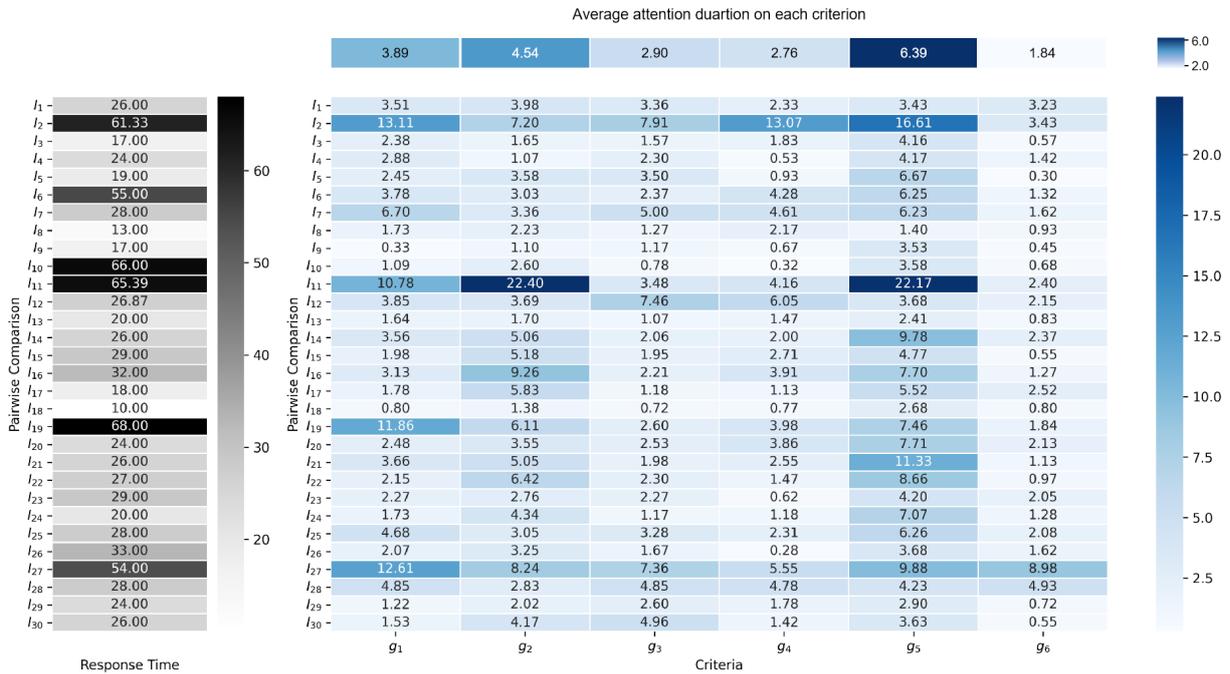

Figure C.3: Response times and attention durations of $d_{29}$ across tasks.